\documentclass[pre,reprint,floatfix,showkeys]{revtex4-2}

\usepackage{graphicx}
\usepackage{dcolumn}
\usepackage{bm}
\usepackage{CJK}
\usepackage{xcolor}
\usepackage{amsmath}
\usepackage{amsfonts}
\usepackage[nomessages]{fp}
\usepackage{appendix}

\usepackage{verbatim}

\begin{document}
\begin{CJK*}{GB}{} 

\title{Nodal statistics-based equivalence relation for graph collections}
\author{Lucrezia Carboni$^{1,2}$ \qquad  {Michel Dojat}$^{2}$ \qquad Sophie Achard$^{1}$ }
\affiliation{$^{1}$ Univ. Grenoble Alpes, CNRS, Inria, Grenoble INP, LJK, 38000 Grenoble, France \\
$^{2}$ Univ. Grenoble Alpes, Inserm, U1216, Grenoble Institut Neurosciences, GIN, 38000 Grenoble, France}

\date{\today}

\begin{abstract}
Node role explainability in complex networks is very difficult, yet is crucial in different application domains such as social science, neurosciences or computer science. Many efforts have been made on the quantification of hubs revealing particular nodes in a network using a given structural property. Yet, in several applications, when multiple instances of networks are available and several structural properties appear to be relevant, the identification of node roles remains largely unexplored. Inspired by the node automorphically equivalence relation, we define an equivalence relation on graph nodes associated with any collection of nodal statistics (i.e. any functions on the node-set). This allows us to define new graph global measures: the power coefficient, and the orthogonality score to evaluate the parsimony and heterogeneity of a given nodal statistics collection. In addition, we introduce a new method based on structural patterns to compare  graphs that have the same vertices set. This method assigns a value to a node to determine its role distinctiveness in a graph family. Extensive numerical results of our method are conducted on both generative graph models and real data concerning human brain functional connectivity. The differences in nodal statistics are shown to be dependent on the underlying graph structure. Comparisons between generative models and real networks combining two different nodal statistics reveal the complexity of human brain functional connectivity with differences at both global and nodal levels. Using a group of 200 healthy controls connectivity networks, our method computes high correspondence scores among the whole population, to detect homotopy, and finally quantify differences between comatose patients and healthy controls.
\keywords{graph comparison; node roles detection; human brain functional connectivity.}
\end{abstract}

\maketitle
\end{CJK*}


\section{\label{sec:INTRO} INTRODUCTION}
In several application scenarios which focus on complex network studies, being able to determine node roles has proven to be relevant \cite{de2014role,weng2007movie,borgatti2009network,finotelli2021graphlet,PhysRevLett.125.248301}. Indeed, the notion of node roles has been introduced in social science structural theory \cite{10.2307/270991} with at least two different conceptions: structural equivalence and structural isomorphism. According to the former, nodes are equivalent if they share exactly the same neighbors. For the latter, nodes are equivalent if there exists an automorphism which maps the first node to the second and \textit{vice versa}. In this work, we consider this latter conception and identify the node role with its structural equivalence class. 

Recently, node roles analysis has been applied to various application domains such as web graphs \cite{meghabghab2002discovering}, technological or biological networks \cite{rossi2014role}. Different algorithms have been proposed to detect structural equivalence classes in a single network by evaluating similarity metrics among nodes \cite{yu2021rolesim,jeh2002simrank,chen2020gaussian}.

When examining a network collection defined on the same node-set, node role detection can provide meaningful information for collection characterization, possibly revealing a specific nodal partitioning.  Indeed, in many real-world applications, the available graph set can potentially be characterized by specific node role classes \cite{KKMMN2016,borgwardt2005protein,cardillo2013emergence,de2015structural,hu2020open}. 
However, while many graph comparison metrics already exist \cite{wills2020metrics,schuld2020measuring}, there is no evidence of a method for comparing them, moreover, none of them directly address the detection of differences at the nodal level. 

This work has been motivated by our interest in human brain functional connectivity networks. In such networks, node organization has proven to be critical, for instance in consciousness states differentiation \cite{achard2012hubs}. However, while current methods allow to discriminate brain networks under various pathological conditions \cite{dadi2019benchmarking,richiardi2013machine}, interpretation and explanation of the exhibited differences between graphs at the nodal level remain difficult. \\
The contributions of this work are then fourfold. First, we define a structural equivalence relation on a graph node-set based on nodal statistics (any functions on the node-set). The proposed definition 
allows determining node role classes according to statistics values. The main innovation of this definition is given by the possibility of identifying the graph structural pattern based on an original combination of as many statistics as desired. 
Second, we define two global measures of a statistics set which determine parsimony and heterogeneity of its elements. These measures only depend on the graph structure and can be used for statistics selection or {graph complexity evaluation \cite{bianconi2007entropy}}.
Third, we propose to compare graphs with the same vertices according to their structural patterns similarity. Indeed, thanks to the identification of node classes, we can compare different graph instances throughout the evaluation of the similarity of their structural patterns. 
Finally, we propose a framework to determine node categories in a network group which allows to characterize the group at a nodal level and to discriminate nodes according to their role. 
     
\begin{figure*}[!ht]
\includegraphics[width=17.2cm]{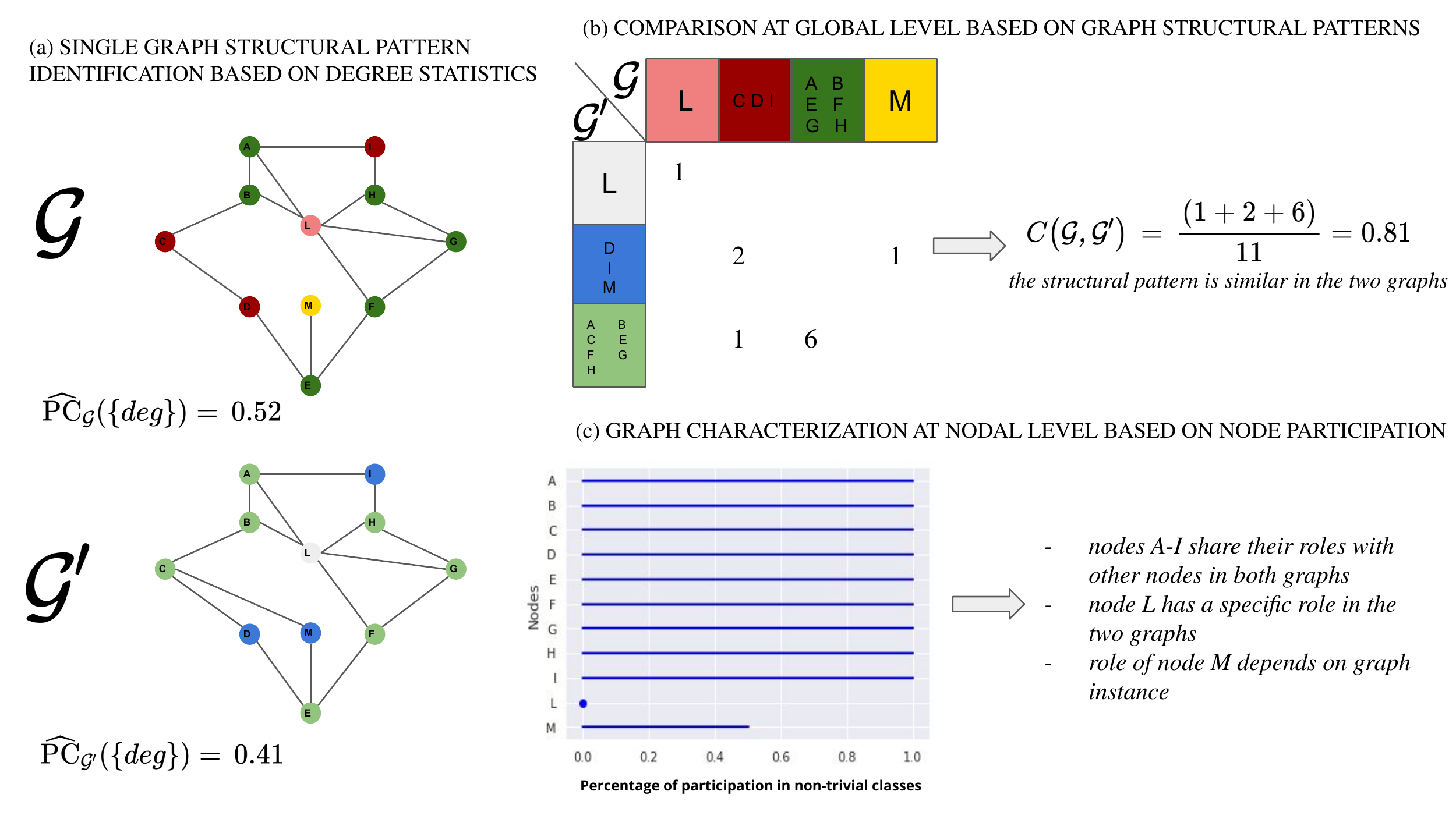}
\caption{\label{fig:framework} Global comparison and nodal characterization of graphs: (a) structural patterns associated with the same statistics are determined on the graphs, (b) the structural patterns are matched to compute a similarity value, (c) nodal participation in non-trivial classes is obtained for nodal characterization.}
\end{figure*}

\section{\label{sec:2} STRUCTURAL EQUIVALENCE FOR A SINGLE UNDIRECTED UNWEIGHTED GRAPH }
We propose to identify the graph structural pattern with the equivalence classes of a  {newly} defined equivalence relation. The traditional definition identifies two nodes as automorphically equivalent if it exists a node permutation preserving the adjacency matrix (an automorphism) which maps the first node to the second and \textit{vice versa}  \cite{everett1994regular}.\\ 
We define the structural equivalence relation as the union of many equivalence relations, each one associated with a single nodal statistics on the graph.  
When we refer to nodal statistics, we consider any function on the node set $s:\mathcal{V}\to s(\mathcal{V})$ which is a function of the adjacency matrix, i.e. node degree, clustering coefficient of a node, {centrality measures,} etc.
We observe that for every pair of automorphically equivalent nodes $u,v\in\mathcal{V}$, any nodal statistics $s$ is preserved. Therefore, we propose to define an equivalence relation  $\sim_s$, associated with a statistics $s$, on the nodes set $\mathcal{V}$ of a graph as follows: \begin{equation}
    v \sim_s u \iff s(u)=s(v).
\end{equation} 

For a nodal statistics having as $s(\mathcal{V})$ a dense and continuous subset of $\mathbb{R}$, the equivalence is defined up to a fixed positive small $\epsilon$: $ v \sim_s u \iff |s(u)-s(v)| \leq \epsilon$ (Supplementary Material (SM) Fig. \ref{fig:epsilon} ).  
As $\sim_s$ is an equivalence relation on $\mathcal{V}$, it is possible to find its induced partition $P$ on $\mathcal{V}$, 
\begin{equation}P_s=\frac{\mathcal{V}}{\sim_s}=\{[a]_{l,{\sim_s}}      \quad\forall l\in s(\mathcal{V})\},\end{equation}
which defines the structural pattern of $\mathcal{G}$ associated with the statistics $s$, and whose elements are the classes of equivalence $[a]_l, \forall l \in s(\mathcal{V})$, \begin{equation}[a]_{l,{\sim_s}}=[a]=\{b\in\mathcal{V} | a \sim_s b \iff s(a)=s(b)=l \}. \end{equation}
A necessary condition for two nodes to be automorphically equivalent is to belong to the same equivalence class. \\
Subsequently, we extend the equivalence relation associated with a statistics to any statistics collection  $\mathcal{S}=\{s_i\}_{i=1,..,n}$, requiring that:
\begin{equation}
a \sim_{\mathcal{S}} b \iff  a \sim_{s_1}b, \, a \sim_{s_2}b,\, \dots, \, a \sim_{s_n}b .    
\end{equation}
Again, we can determine {$P_\mathcal{S}=\{ [a]_{\sim_{\mathcal{S}}} \}$} the induced partition by $\sim_{\mathcal{S}}$ on $\mathcal{V}$ as the intersection of each class of the considered $\{s_i\}_{i=1,..,n}$.
A visualization of the partitions associated with degree statistics is shown in Fig. \ref{fig:framework} (a).

Since the automorphically equivalence relation preserves any nodal statistics, the nodal statistics-based equivalence relation associated with an infinity collection retrieves the automorphically equivalence. However, a finite nodal statistics collection with this property may also exist. 
We propose to compare statistics collection according to new defined global graph parameters which measure respectively parsimony and heterogeneity of its elements. These global parameters depend on the graph structure.\\
{Given $P_\mathcal{S}$, one can compute exactly the number of eligible automorphisms that map nodes into the
same equivalence class, as it is computed below.}
Therefore, for each statistics collection on a graph ${\mathcal{G}}$, we can estimate how many permutations are prevented from being tested as being adjacency preserving in a brute force approach. We introduce the power coefficient (PC) of a set $\mathcal{S}$ for a graph ${\mathcal{G}}=(\mathcal{G},\mathcal{E})$
\begin{equation} \text{PC}_{\mathcal{G}}(\mathcal{S}) =  { \Big | \log \frac{\#\{\text{permutations preserving } P_{\mathcal{S}}\}}{\#\{\text{permutations of }\mathcal{V}\}} \Big | } \end{equation}
with 
\begin{align*}
&\#\{\text{permutations preserving } P_{\mathcal{S}}\} = \prod_{o\in P_{\mathcal{S}}}|o|! \\
&\#\{\text{permutations of }\mathcal{V}\} = |\mathcal{V}|! .
\end{align*}
The value {$|\mathcal{V}|!e^{-\text{PC}}$} corresponds to an upper bound of the number of automorphisms of $\mathcal{G}$. Indeed, PC is increasing when more nodal statistics are combined together (SM, Fig. \ref{fig:pc}). 
{In the special case in which the permutations preserving $P_{\mathcal{S}}$ can be identified with the automorphisms of $\mathcal{G}$, PC can be interpreted as entropy of the network ensemble \cite{bianconi2007entropy} having $\mathcal{G}$ topology (SM, \ref{sec:SuppMat}). In all other cases, PC encodes the amount of information given by $\mathcal{S}$ on the structure of $\mathcal{G}$ and it is a parsimony measure for $\mathcal{S}$. \\
Since PC takes values in $[0,\log \frac{1}{|\mathcal{V}|!}[$, with an upper bound strictly depending on the number of nodes, we propose a normalized version of PC, $\widehat{\text{PC}}\in [0,1]$ :
\begin{eqnarray}
    \widehat{\text{PC}}_{\mathcal{G}}(\mathcal{S}) &=& \frac{\text{PC}_{\mathcal{G}}(\mathcal{S})}{\log |\mathcal{V}|!} \\
    &=& 1 - \frac{\log \#\{\text{permutations preserving } P_{\mathcal{S}}\} }{\log\#\{\text{permutations of }\mathcal{V}\}} 
\end{eqnarray}
The higher the $\widehat{\text{PC}}$, the more the collection of statistics $\mathcal{S}$ capture the heterogeneity of nodal structural roles in $\mathcal{G}$. Indeed, for a vertex-transitive graphs (i.e. all nodes are automorphically equivalent) $\widehat{\text{PC}}_{\mathcal{G}}(\mathcal{S})=0 $ for all nodal statistics $\mathcal{S}$, while if it exists a collection $\Bar{\mathcal{S}}$ s.t. $\widehat{\text{PC}}_{\mathcal{G}}(\Bar{\mathcal{S}})= 1$ then the graph $\mathcal{G}$ does admit non-trivial automorphisms. 
\\}
Hence, we introduce the notion of perfectly orthogonal statistics for heterogeneity evaluation of a collection elements. First, two nodal statistics are said to be perfectly orthogonal if their union-associated equivalent relation induces the trivial partition: all nodes belong to a single element set. Next, we extend the definition to any nodal statistics set: a nodal statistics collection is said to be perfectly orthogonal if its induced partition is trivial.
An orthogonality measure for a given nodal statistics set on a graph can be assessed by computing the number of nodes in non-trivial classes on its associated partition:
\begin{equation} O_{\mathcal{G}}(\mathcal{S})=\frac{|\{v \in \mathcal{V} \, \text{s.t.} \, \#[v]_{\sim_{\mathcal{S}}} \neq 1\}|} {|\mathcal{V}|}\end{equation}
$O_{\mathcal{G}}(\mathcal{S})$ is the ratio between the number of nodes in non-trivial classes and the total number of vertices and corresponds to an orthogonality score. By definition, $\mathcal{S}$ is perfectly orthogonal if and only if $O_{\mathcal{G}}(\mathcal{S})=0$. 

\section{\label{sec:3} STRUCTURAL EQUIVALENCE FOR  GRAPH COLLECTIONS}
\subsection{Structural pattern comparison}
\begin{figure}[!htp]
\begin{minipage}{8.6cm}
\includegraphics[trim={1cm 3cm 1cm 0 },clip,width=8.6cm]{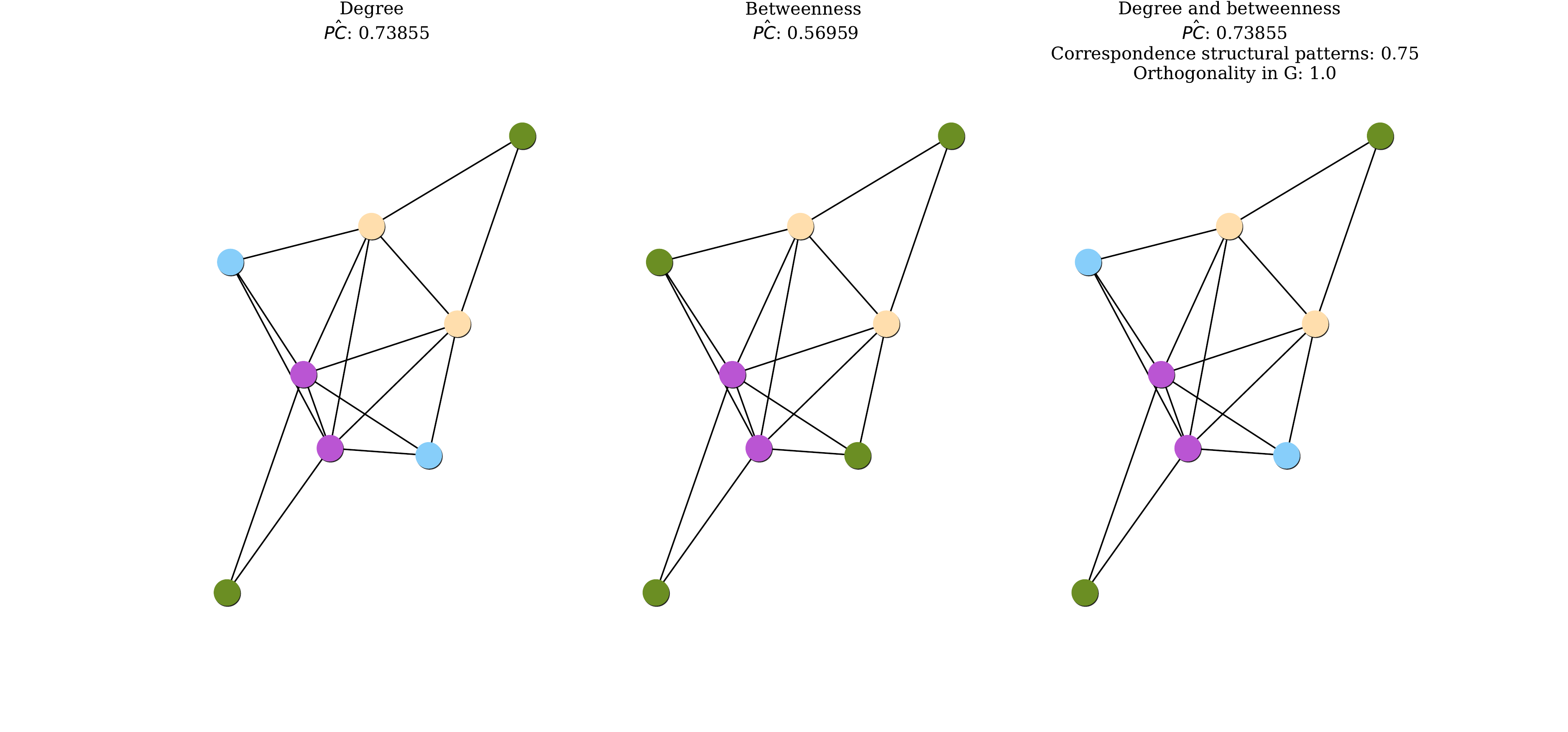}
{(a) One statistics is enough informative}
\end{minipage}
\hfill
\begin{minipage}{8.6cm}
\centering
\includegraphics[trim={1cm 3cm 1cm 0 },clip,width=8.6cm]{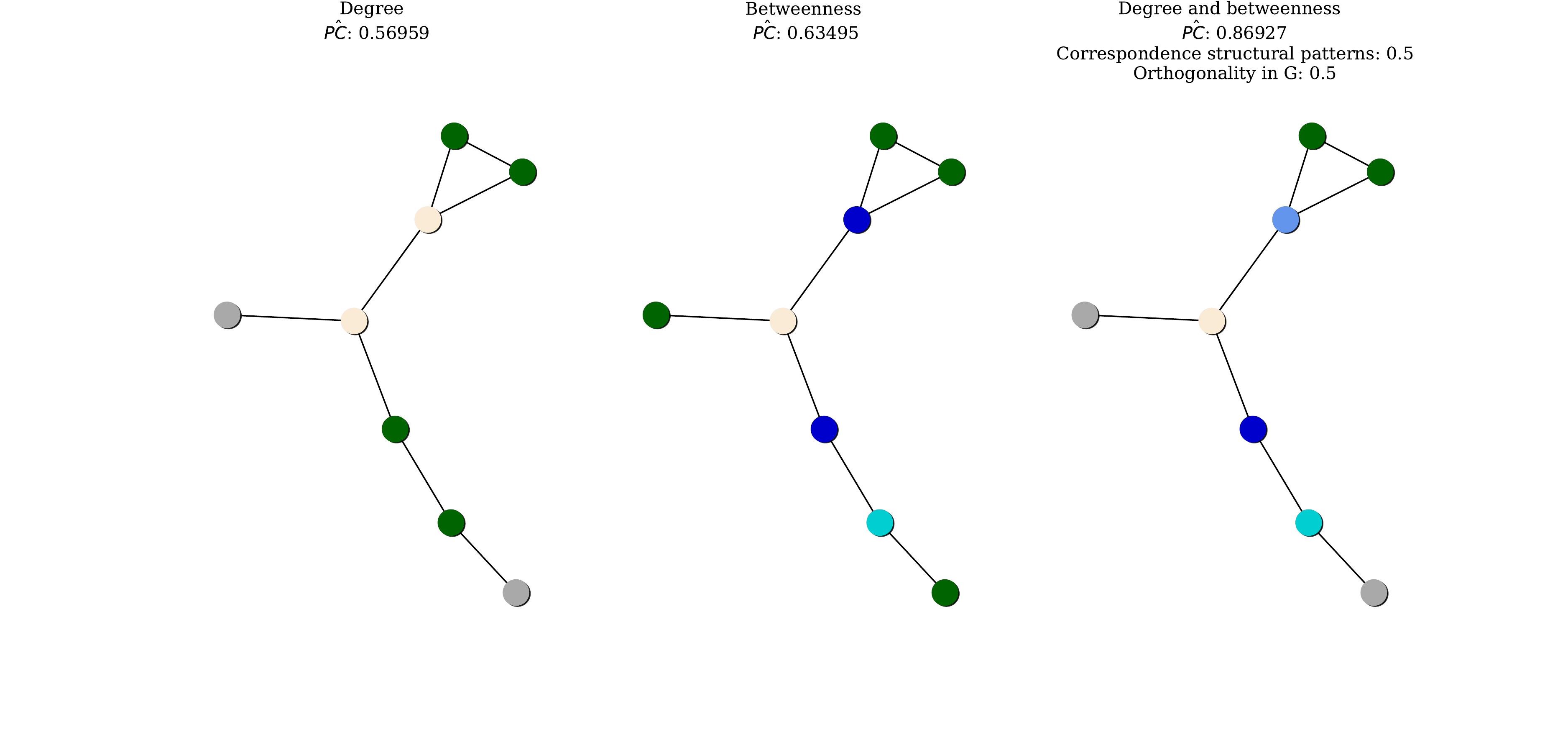}
(b) Two nodal statistics are more informative 
\end{minipage}
\hfill
\begin{minipage}{8.6cm}
\includegraphics[trim={1cm 0 1cm 0 },clip,width=8.6cm]{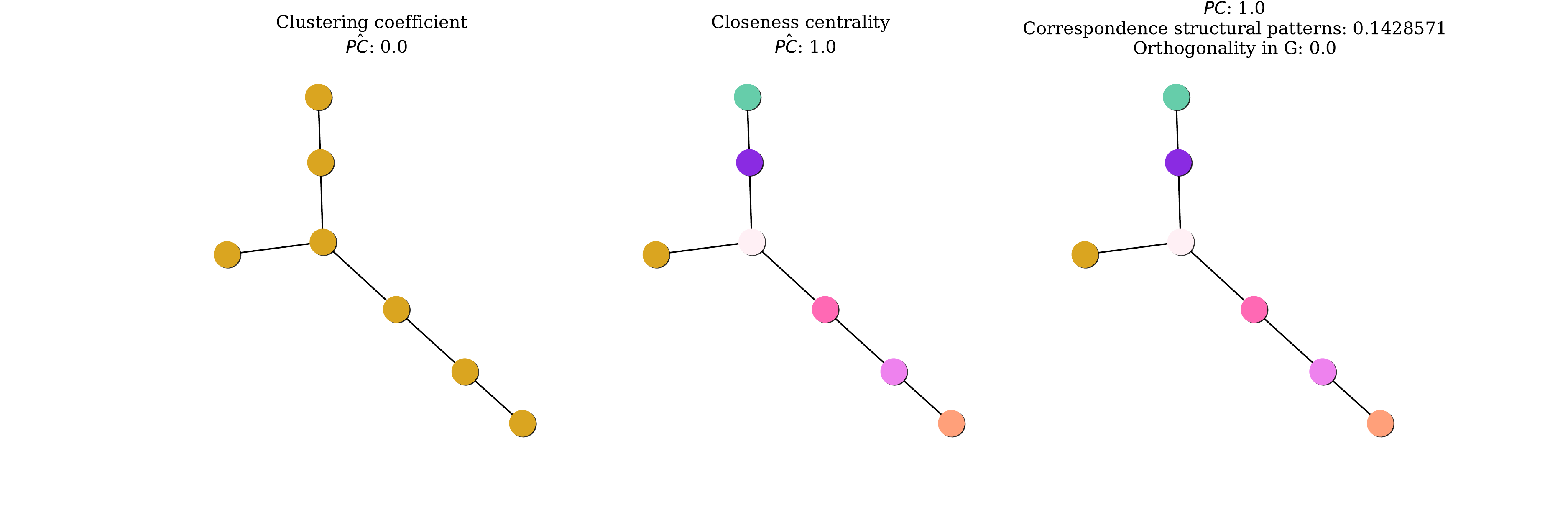}
{(c) Two perfectly orthogonal nodal statistics}
\end{minipage}
\caption{\label{fig:examples} Visualization of global metrics on three graphs (column) for three cases (row).  (a) One statistics is sufficient: the $\widehat{\text{PC}}$ value observed when considering two statistics (right) is equal to degree alone (left), meaning that betweenness provides with no useful information for determining the structural pattern. When we compare the two partitions with the degree  statistics alone, we observe $75\%$ of the nodes belonging to the same class and no trivial class (orthogonality equals to 1). This result can be interpreted in two different ways: more node statistics are needed to identify hub nodes, or the considered graph does not contain hub nodes. (b) Two statistics are more informative: to associate degree and betweenness improves the power coefficient. The identified patterns share half of nodes and their orthogonality is 0.5, meaning that their partition 
situates half of nodes in trivial classes. (c) Perfect orthogonality: The minimal values is reached when one of the two compared structural pattern has only one class and the other contains trivial classes.}
\end{figure}

Graphs that have the same node set can be compared by evaluating the correspondence between their structural patterns.
The node set constraint can be easily circumvented when two graphs do not share all the nodes, by including all nodes to the graphs vertices set and allowing the network to be composed of more connected components. Indeed, each network can be seen as the union of one strongly-connected component with as many single disconnected vertices as needed. \\
We propose to compare structural patterns as follows. Let $\mathcal{G},\mathcal{G'}$ be two graphs having same vertices $\mathcal{V}$ and let $\mathcal{S}$ be a statistics collection whose associated partitions are $P_{\mathcal{S}}, P'_{\mathcal{S}}$ on $\mathcal{G},\mathcal{G}'$ respectively. 
Given bijective mapping from $P_{\mathcal{S}},P'_  {\mathcal{S}}$ to an initial segment of the natural numbers as enumerations, let $c(v_i),c'(v_i)$ be the enumeration of the classes of $v_i$, the correspondence structural pattern score between $\mathcal{G},\mathcal{G'}$  is defined as:
\begin{equation} C(\mathcal{G},\mathcal{G'})= \max_{\pi\in\Pi}\frac{1}{|\mathcal{V}|} \sum_{i=1}^{|\mathcal{V}|} \mathcal{X}(\pi(c(v_i))=c'(v_i))
\end{equation}
where $\Pi$ is the set of all coupling between the elements in $P_{\mathcal{S}}$ and the elements in $P'_{\mathcal{S}}$ and $\mathcal{X}$ is the indicator function. \\
{A possible implementation of $C(\mathcal{G},\mathcal{G'})$ in polynomial time is given by the Hungarian algorithm (\cite{kuhn1955hungarian}) for assignment problems with has a complexity $\mathcal{O}(\max\{|P_{\mathcal{S}}|,|P'_{\mathcal{S}}|\}^3 )$ which in the worst case equals $\mathcal{O}(|\mathcal{V}|^3)$. \\}
The correspondence structural pattern score can be applied for two different purposes: to evaluate structural pattern similarity between two graphs (Fig. \ref{fig:framework} (b)) or to evaluate the similarity of structural patterns associated with different statistics collection on the same graph. 
Since at least one class of $P_{\mathcal{S}}$ shares one element with one of the classes in $P'_{\mathcal{S}}$, $C(\mathcal{G},\mathcal{G'})\geq\frac{1}{|\mathcal{V}|}$. As a consequence,  even perfectly orthogonal statistics set of a graph can exhibit a correspondence pattern score higher than zero (SM Fig. \ref{fig:examples} (c)). \\

If for every class in $P_{\mathcal{S}}$ there exists one class of $P'_{\mathcal{S}}$ having all and only its elements, then $P_{\mathcal{S}} = P'_{\mathcal{S}}$ and $C(\mathcal{G},\mathcal{G'}) = 1$. The opposite is also true: same partitions determine a correspondence structural pattern score equals to 1.

More general properties of the defined global measures can be found in the SM \ref{sec:SuppMat}.

\subsection{Nodes distinctiveness or similarity}
Since eligible automorphisms can only map nodes within classes, a node in a trivial class (one element class) is always a fixed point. Thus, to provide a group characterization at nodal level, we propose to enumerate for each node its participation into non-trivial classes as a measure of the node's propensity not to be a fixed-point of admissible automorphisms.
The more a node appears in non-trivial classes, the more it shows common properties with some other nodes in the graph. The persistence of a node to belong to a class in an entire graph group reveals the presence of shared properties among the group for the given node, i.e. hubs nodes, peripheral nodes, etc (Fig. \ref{fig:framework} (c)). Thus, given a graphs group $ G=\{\mathcal{G}_k=(\mathcal{V}_k,\mathcal{E}_k) \, \,  \text{s.t.}\,\, \mathcal{V}_k= \mathcal{V}\}$, {and a statistics collection $\mathcal{S}$} we count the percentage of participation of each node of $\mathcal{V}$ in non-trivial classes:
\begin{equation}
\forall v\in \mathcal{V} \quad {\text{PP}^{\mathcal{S}}_{G}(v)} = \text{PP}_{G}(v) = \frac{|\{\mathcal{G}_k \in G \,\, \text{s.t.} \,\, \#{[v]^{\mathcal{G}_k}_{\sim_\mathcal{S}}} \neq 1 \}|}{|G|} \end{equation}
with {$[v]^{\mathcal{G}_k}_{\sim_{\mathcal{S}}}$ the class of $v$ in $\mathcal{G}_k$ in the partition induced by $\mathcal{S}$. In the following, with abuse of notation, we suppose $\mathcal{S}$ fixed and avoid to explicitly repeat the dependency.} A high percentage of participation means the node shares its role in many graph instances in the group, while at the opposite a node which does not share its role consistently shows a distinctiveness behavior in the considered graphs collection.  \\

\section{\label{sec:4} EXPERIMENTS}
\subsection{Synthetic data} 
We consider different generative graph models and compare them according to their sparsity level, defined as the ratio between the edge count on the graph and the edge count in a complete graph having the same nodes. 
We fix the number of nodes to 90 to be in line with the considered real dataset. Indeed, 90 corresponds to the number of human brain regions classically used in brain partitioning \cite{tzourio2002automated}. We examine Erd\H{o}s-R\'enyi \cite{Erdos:1959:pmd} (ER), Watts-Strogatz \cite{watts1998collective} (WS) and Barab\'asi-Albert \cite{barabasi1999emergence} models (BA1, BA2). Moreover, to be close to real situations, we consider additional models driven by human brain data. Here, such models provide synthetic versions of corresponding real networks: a model which preserves the degree sequence (DSP), two models of brain connectivity, economical preferential attachment (EPA) and economical clustering (EC), models proposed in \cite{vertes2012simple}. More models details are provided in SM. \\
In our experiments, we consider classical graph statistics: degree, clustering coefficient, and centrality measures (betweenness, closeness, second-order) \cite{newman2005measure,brands2001faster,PhysRevE.76.026107,PhysRevE.71.065103,katz1953new,freeman1979centrality,kermarrec2011second}. 

\subsection{Real data: Human brain functional connectivity networks} 
Our framework has been developed to provide new statistical tools for the quantitative analysis and comparison of brain functional connectivity networks. To give a flavour of this application, we consider 200 networks built from resting-state functional magnetic resonance imaging (RS-fMRI) available through Human Connectome Project (HCP) \cite{van2012human,termenon2016reliability}. The brain was parcelled in ninety regions (AAL90 atlas) \cite{tzourio2002automated}.  For each region, a unique time-series signal was determined by averaging the RS-fMRI time-series over all voxels, weighted by the gray matter proportion. Then, wavelet correlation \cite{bullmore2004wavelets} among regional time-series was estimated at 0.043-0.087 Hz frequency interval \cite{biswal1995functional,lowe1998functional,cordes2000mapping,salvador2005undirected}. Finally, the correlation matrices were thresholded to extract graphs at various sparsity ratios \cite{de2014graph,achard2006resilient}. 
When analyzing graph at fixed sparsity, we select 0.1 which guarantees that each extracted network belongs to small-world regime corresponding to global and local efficiencies comprised between the ones of ER graph and ones of the complete graph \cite{achard2007efficiency,latora2001efficient}. \\

\section{\label{sec:5} RESULTS}
\begin{figure}[!hb]
\includegraphics[trim={2.5cm 1cm 4.5cm 3cm },clip,width=8.6cm]{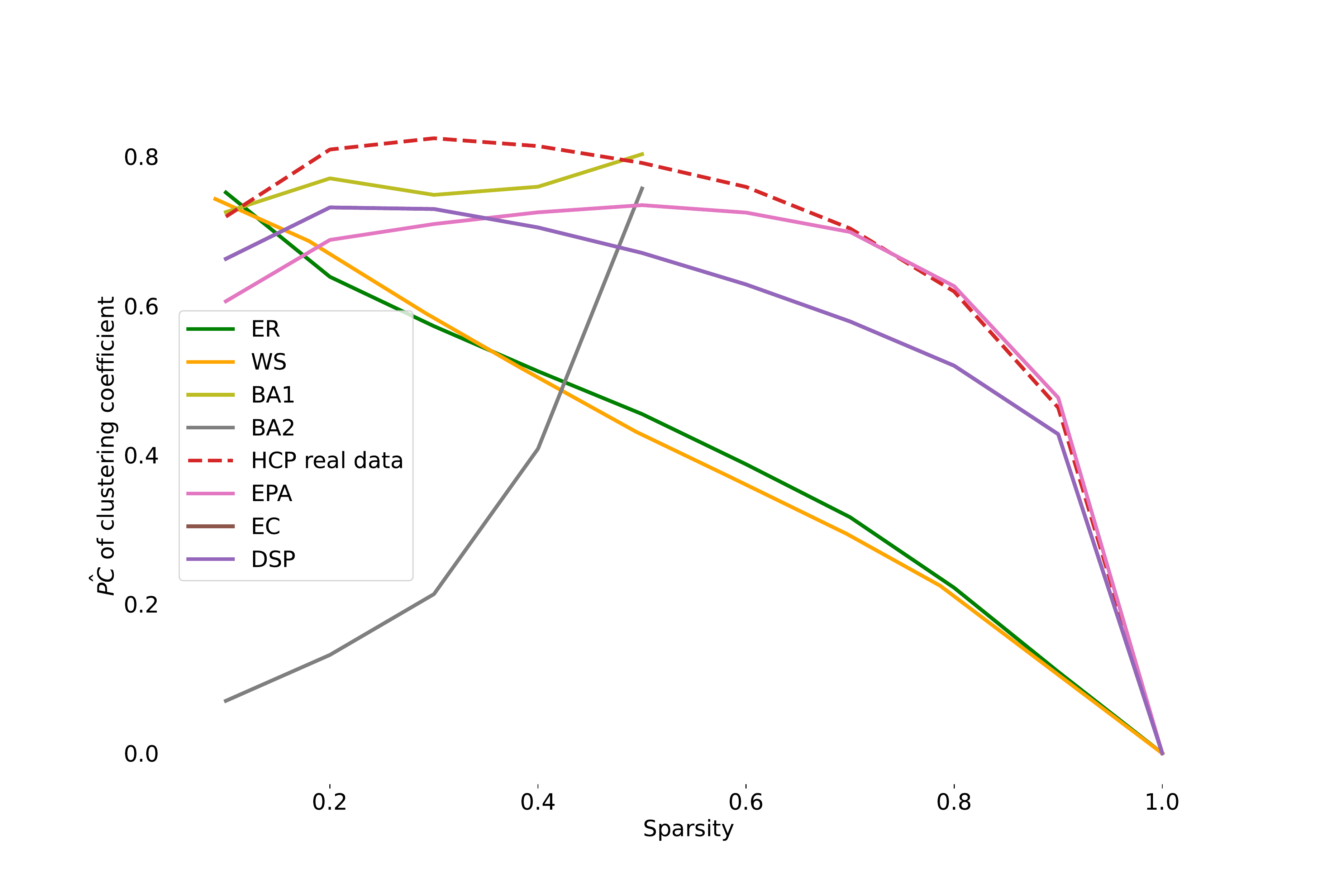}
\caption{\label{fig:pc_deg} Mean normalized power coefficient ($\widehat{\text{PC}}$) of clustering coefficient statistics on different generative models and real brain connectivity graphs (HCP) at different sparsity levels.
ER: Erd\H{o}s-R\'enyi;  WS: Watts-Strogatz; BA1, BA2: Barab\'asi-Albert; DSP: Degree sequence preserving model; EPA: Economical preferential attachment model; EC: Economical clustering model.}
\end{figure}
\subsection{Generative networks}
\begin{figure*}[ht]
\includegraphics[trim=6cm 3.5cm 5.4cm 2cm,clip,width=17.2cm]{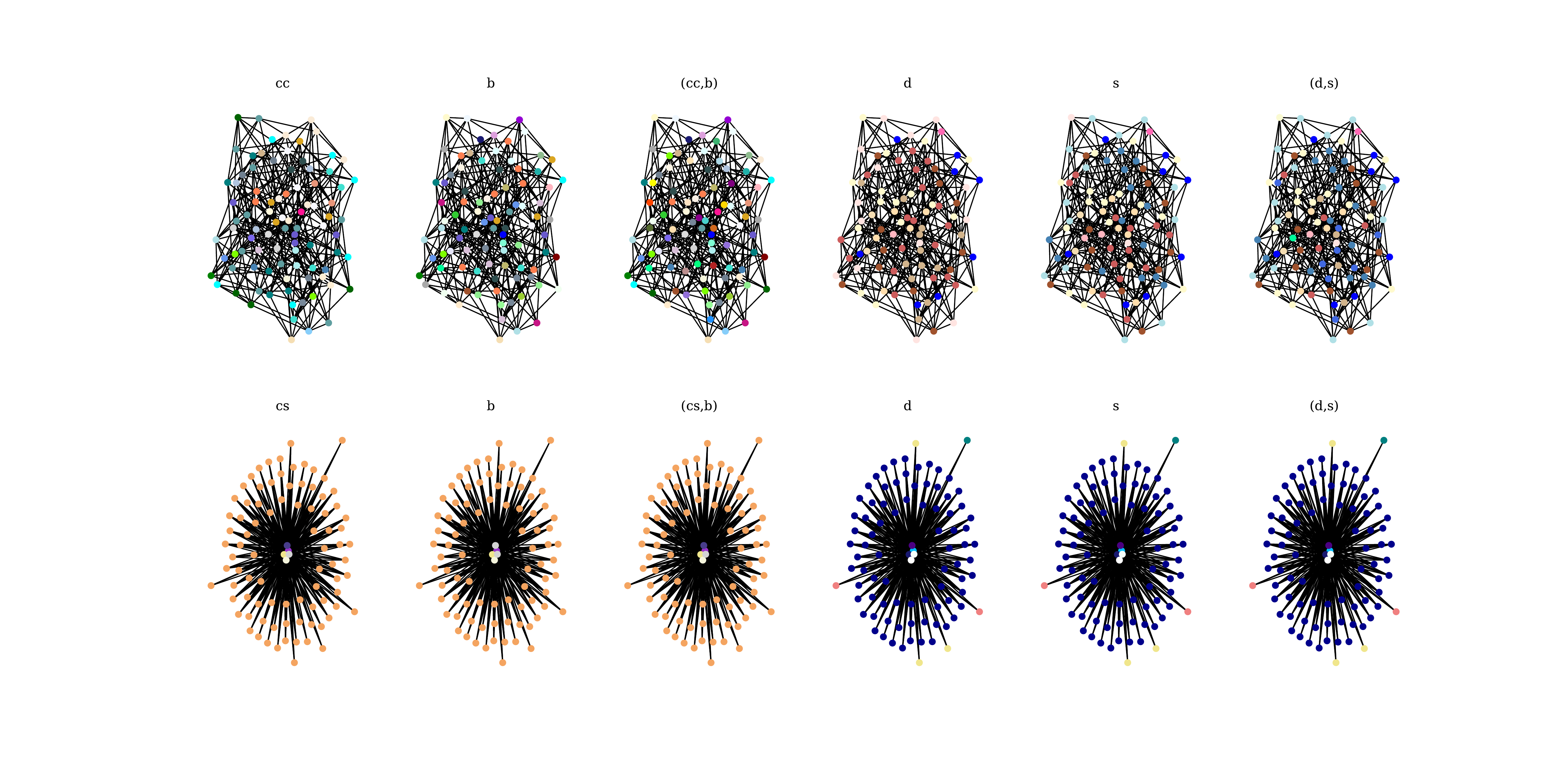}
\caption{\label{fig:structural_visualization} Structural patterns associated with different statistics on Watt-Strogatz (top) and Barab\'asi-Albert (bottom) models. cc: Clustering coefficient; b: Beetweeness centrality; d: degree; s: Second-order centrality; cs: Closeness centrality}
\end{figure*}
\begin{figure}[!ht]
\includegraphics[width=8.6cm]{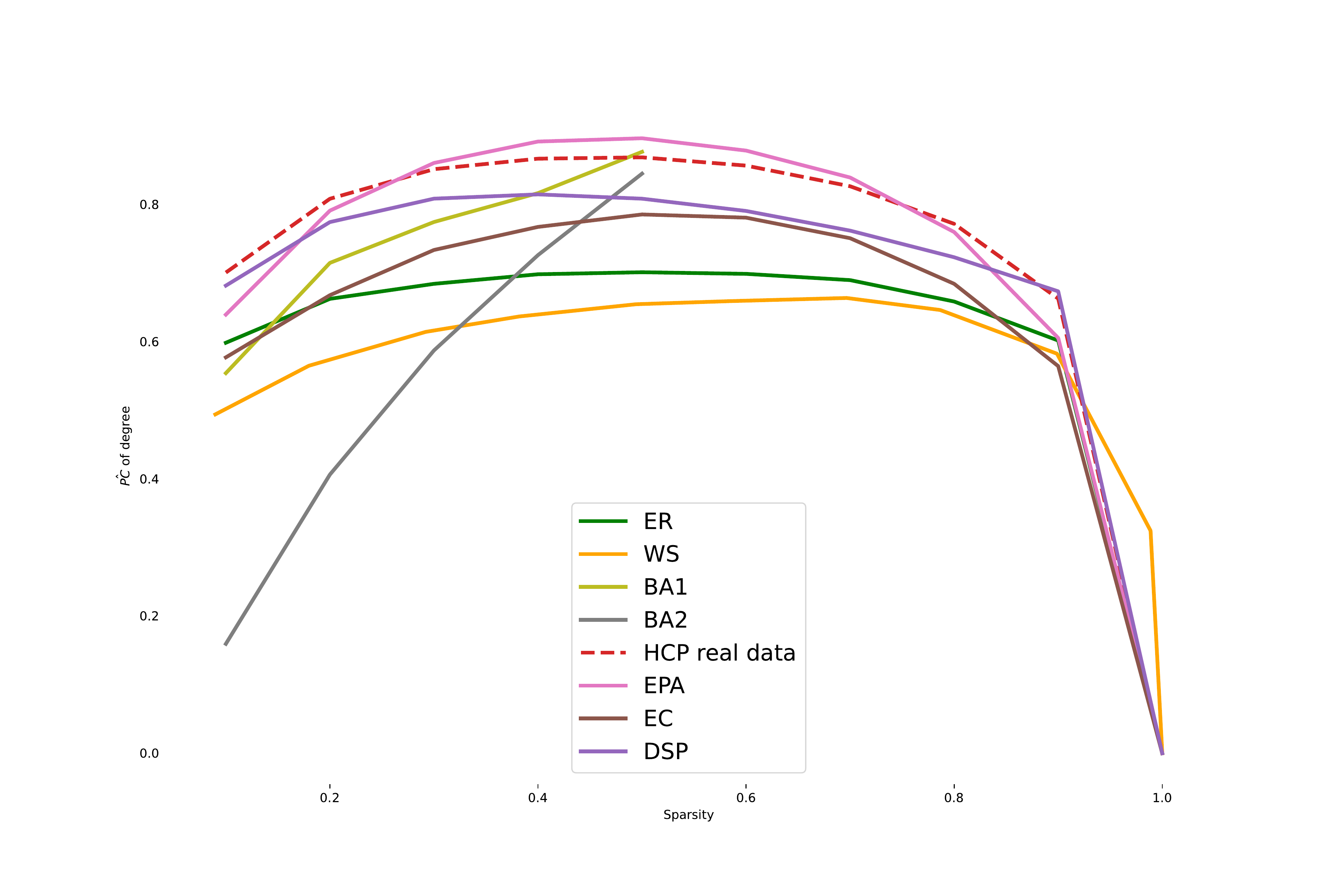}
\caption{\label{fig:pc_degrees} Mean normalized power coefficient ($\widehat{\text{PC}}$) of degree statistics on different generative models and real brain connectivity graphs (HCP) at different sparsity levels. ER: Erd\H{o}s-R\'enyi;  WS: Watts-Strogatz; BA1, BA2: Barab\'asi-Albert; DSP: Degree sequence preserving model; EPA: Economical preferential attachment model; EC: Economical clustering model. Interesting, the $\widehat{\text{PC}}$ on the real data have the best performance at all sparsity levels. When evaluating the $\widehat{\text{PC}}$ of different measures on the same model, we can select for each sparsity ratio which nodal statistics have the higher discriminative power on the node set.}
\end{figure}

In Fig. \ref{fig:pc_deg}, we report clustering coefficient $\widehat{\text{PC}}$ on different generative models and real connectivity graphs with respect to their sparsity ratio. {The same analysis can be conducted for different nodal statistics (SM \ref{fig:pc_degrees}), yet for clustering coefficient, we can easily observe that} for the same level of sparsity, the $\widehat{\text{PC}}$ behaves differently across the models. 
As expected, $\widehat{\text{PC}}=0$ when sparsity ratio $=1$ as in a complete graph, it is not possible to extract any meaningful class for any nodal statistics. \\
For Barab\'asi-Albert model (BA1, BA2), we observe a monotone increasing $\widehat{\text{PC}}$ with respect to the sparsity. Indeed, when the sparsity is low, we have few nodes of high clustering coefficient and many nodes of very low coefficient values. The number of automorphisms exchanging nodes of low values is then higher for small sparsity, while when the sparsity ratio increases the clustering coefficient values distribution tends to be less concentrated on the node set, identifying more classes and corresponding to higher $\widehat{\text{PC}}$. ER and WS show similar behavior especially for high sparsity values, while when the sparsity is low, WS tends to differ from ER model.\\

Regarding brain models, EPA fits correctly the HCP networks when the sparsity is higher than $0.7$. EC and DSP curves follow the HCP curve tendencies, but with lower $\widehat{\text{PC}}$ values. 
A possible explanation of this difference, is the presence of hubs in HCP networks not present in the models. Indeed, a higher number of hubs results in higher $\widehat{\text{PC}}$ score. \\
Then, we evaluate orthogonality and correspondence structural pattern of statistics pairs in WS and BA2 models at 0.1 sparsity (SM Fig. \ref{fig:corr}). A visualization of their structural patterns is shown in Fig. \ref{fig:structural_visualization}.
In WS model, the degree shows high orthogonality values with all nodal statistics: many nodes that have same degree also share a second nodal statistics value. This is likely due to its degree distribution. Indeed, in a general WS graph $\mathcal{G}_{n,k,p}$ all nodes have approximately the same degree $k$ \cite{albert2002statistical}. \\
Thus, there is high chance of retrieving high populated classes associated with degree. Interestingly, the correspondence patterns scores between the degree and the other statistics are low except for the second order centrality (Fig. \ref{fig:structural_visualization} Top Right). Degree and second order centrality capture different topological graph features \cite{kermarrec2011second} and usually appear unrelated in complex networks. However, in the considered case, their induced partitions on the graph largely overlap. Indeed, they exhibit a strong negative correlation coefficient ($-0.98$ in average). Their high orthogonality and high correspondence scores reveal this correlation. \\

In WS model, the statistics pair, which shows the least correspondence pattern scores, is composed by degree and betweenness centrality: while trivial degree classes capture high connected nodes,  
the betweenness centrality better refines the class associated with the average degree value. \\
Completely different results are observed in BA2 model, for which the orthogonality of all considered statistics pair together with their correspondence scores appear close to $1.0$. This shows how in preferential attachment model all statistics pairs determine almost the same structural patterns: few populated classes of high connected nodes and high populated class for the leaves. 
Indeed, for BA model a very high correspondence of structural patterns associated with single statistics is detected (Fig.  \ref{fig:structural_visualization} bottom).

\subsection{Human brain functional connectivity networks} 
\begin{figure}[!ht]
\includegraphics[trim={4.8cm 1.63cm 4.9cm 3.3cm },clip,width=8.6cm]{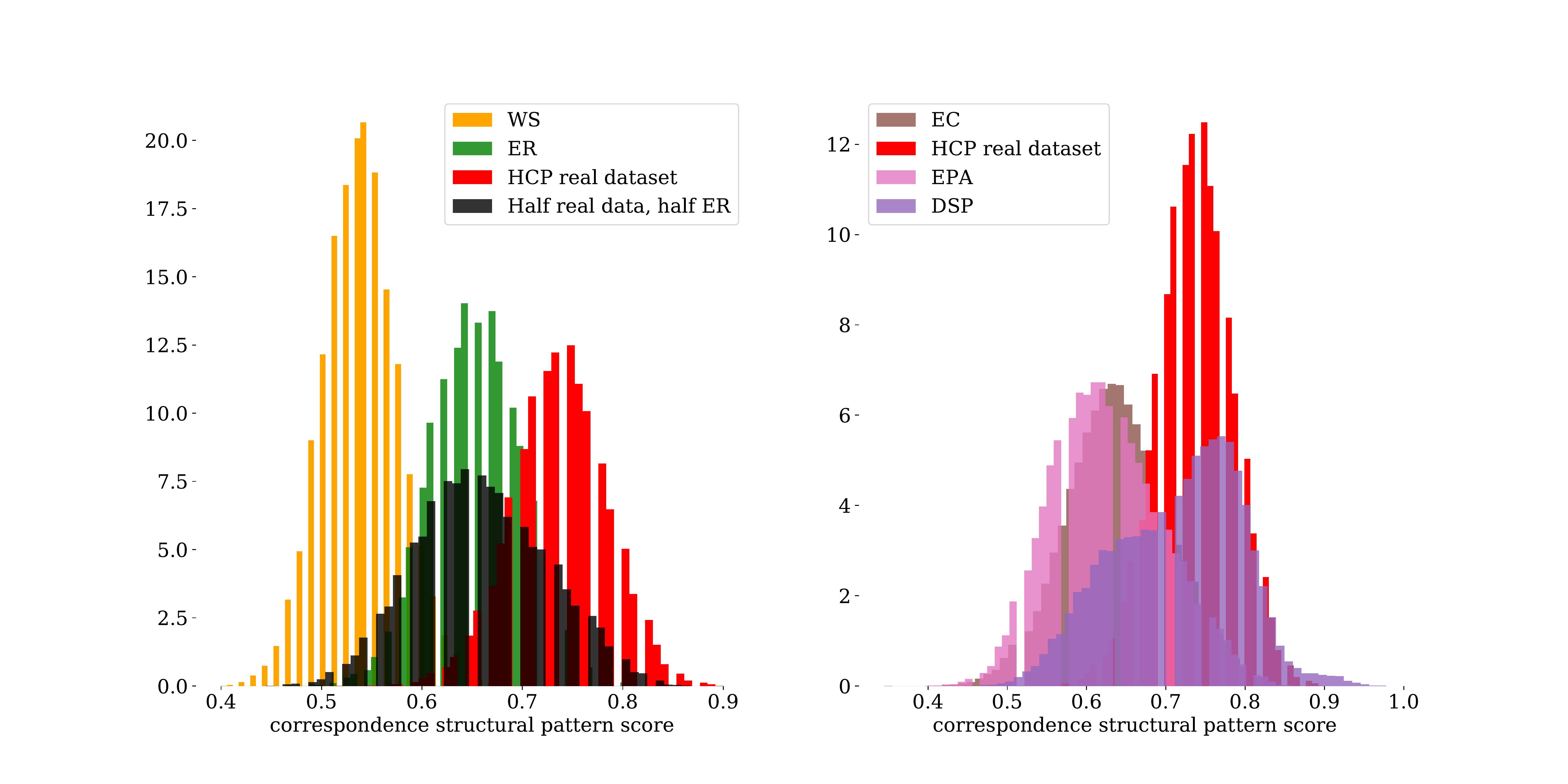}
\caption{\label{fig:orbit_comparison_values} Correspondence structural pattern distribution on the considered model and real data. Left: WS: Watt-Strogatz model, ER: Erd\H{o}s-R\'enyi model and HCP; dataset  100 samples randomly chosen from HCP dataset and from ER model. Right: Brain models and HCP data. DSP: Degree sequence preserving model; EPA: Economical preferential attachment model; EC: Economical clustering model.}
\end{figure}

\begin{figure*}[!ht]
\includegraphics[trim={0 0 0 0.8cm },clip,width=17.2cm]{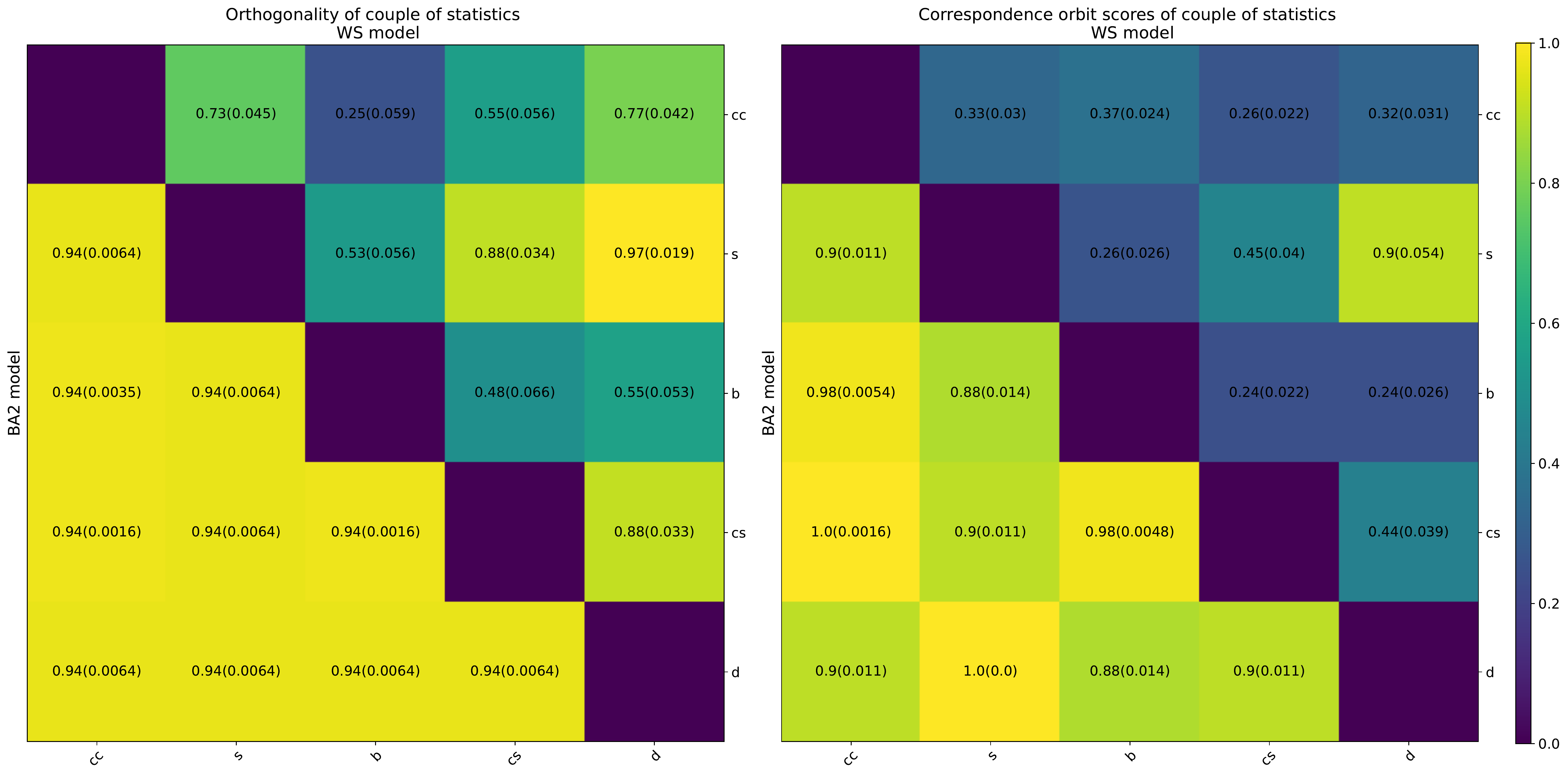}
\caption{\label{fig:corr} 
Nodal statistics pair comparison on two models. Upper Triangular Matrix: WS (Watt-Strogatz model); Lower Triangular Matrix: BA2 (Barab\'asi-Albert) model. Left: Orthogonality for a pair of statistics; Right: Correspondence structural pattern score for a pair of statistics. cc: Clustering coefficient; b: Beetweeness centrality; d: degree; s: Second-order centrality; cs: Closeness centrality}
\includegraphics[trim={0 0cm 0 3cm},clip,width=17.2cm]{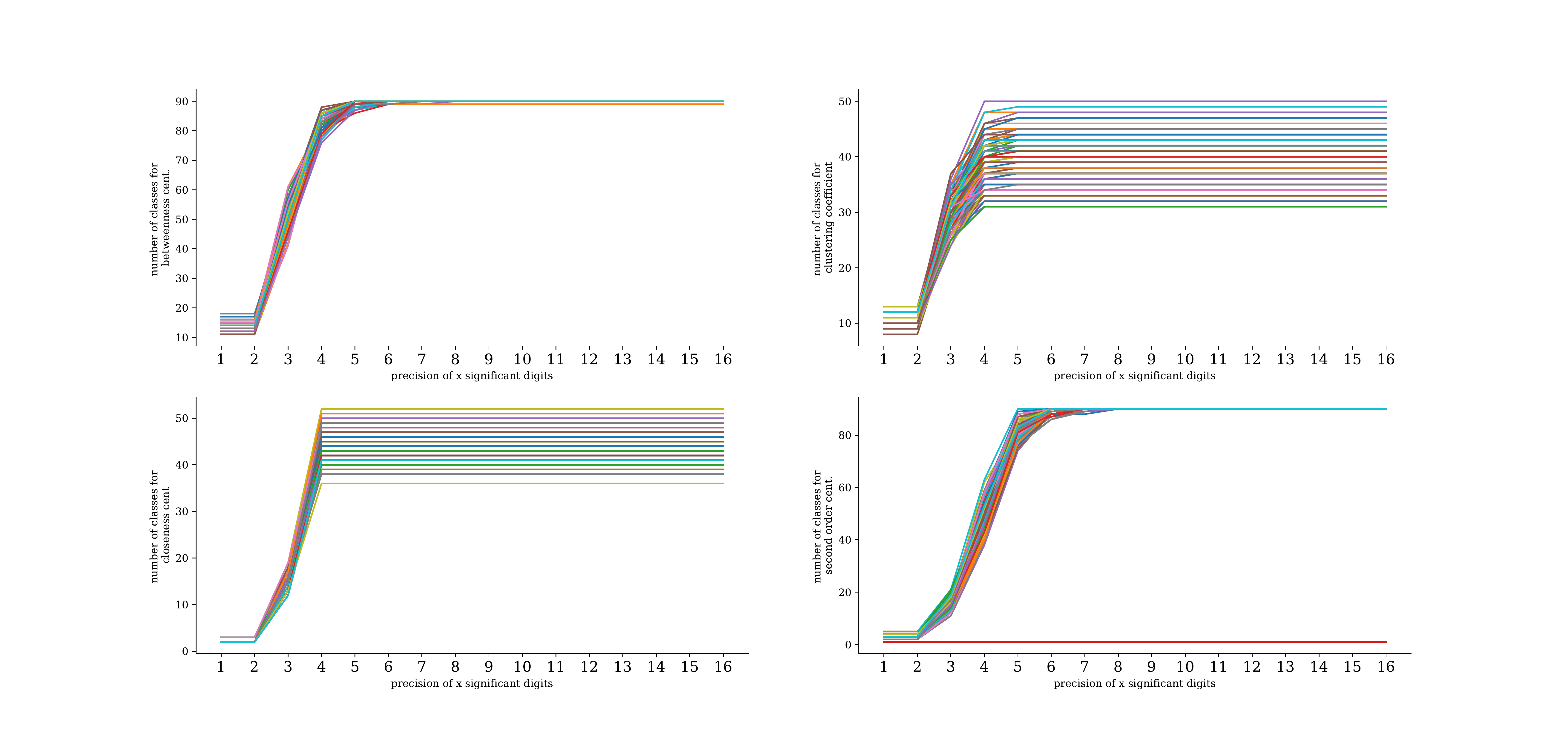}
\caption{\label{fig:epsilon} 
$\epsilon$ choice effects in Erd\H{o}s-R\'enyi graphs. The chosen $\epsilon$ corresponds to a number of significant digits to be used when comparing different nodal-statistics values. When the number of significant digits is higher the number of extracted classes increases. Depending on the considered statistics, the number of classes usually stabilizes around $4$ or $5$ significant digits. Thus, in the experiments $\epsilon$ is fixed to be the minimum number at which the class number stabilizes.}
\end{figure*}

\begin{figure}[!htp]
    \centering
    \includegraphics[width=8.6cm]{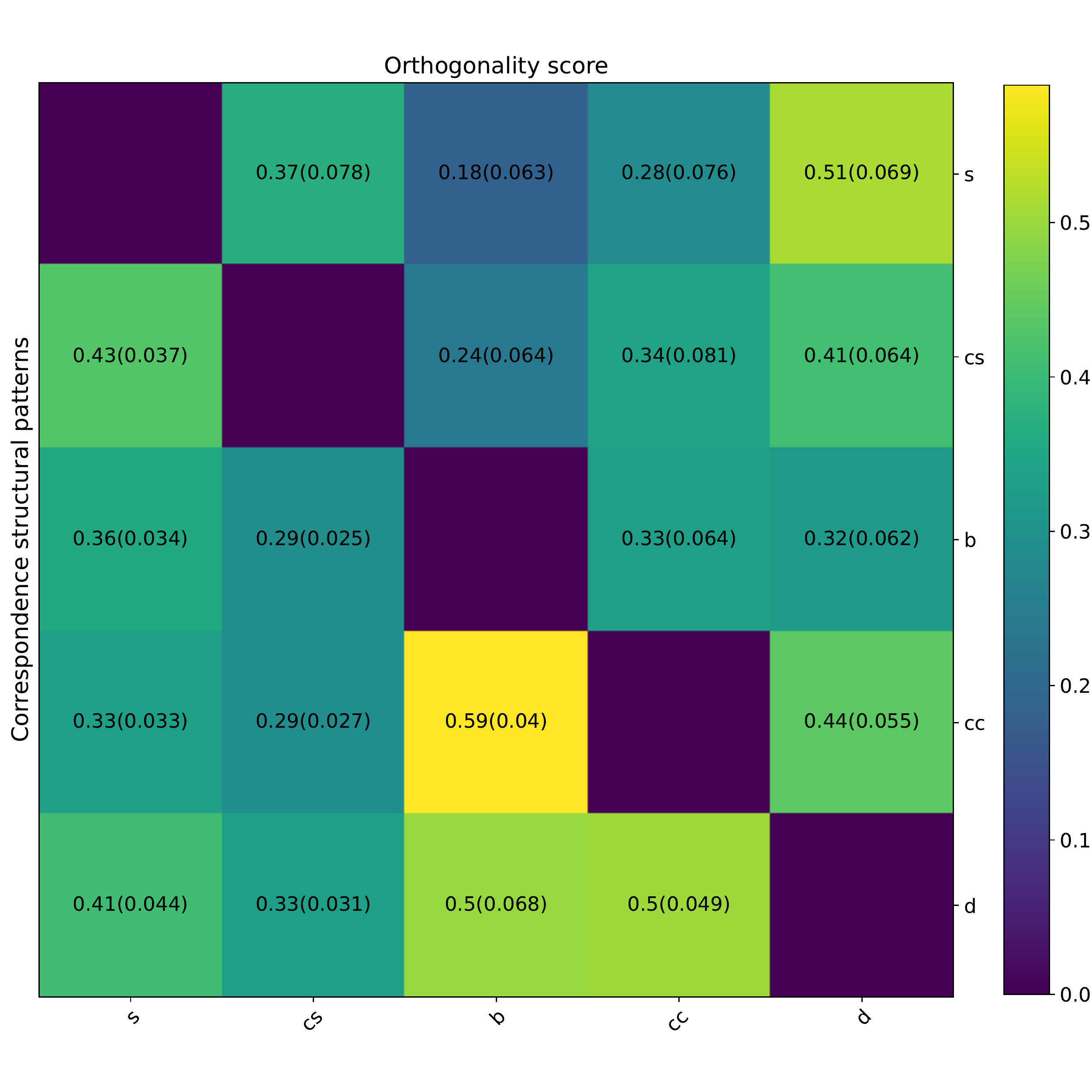}
    \caption{Average value of orthogonality and correspondence structural pattern scores (SD) of nodal statistics pairs on HCP dataset. Upper Diagonal: Orthogonality score, Lower Diagonal: Correspondence structural patterns score. CC: Clustering coefficient; B: Beetweeness Centrality; D: degree; S: Second-order centrality CS: Closeness centrality.}
    \label{fig:ortho_hcp}
\end{figure}

The HCP dataset was analyzed considering degree and betweenness centrality associate-equivalence relation. For this pair, low orthogonality and correspondence patterns scores are observed both on WS model and real data (SM Fig. \ref{fig:corr}, Fig. \ref{fig:ortho_hcp}).\\
We compare the correspondence structural pattern score distribution for generative models and HCP datasets (Fig. \ref{fig:orbit_comparison_values}). The observed ER and WS distribution values are lower compared to real data. Moreover, when considering a dataset combining half HCP real networks and half ER networks, we observe a reduction in the structural pattern comparison values and an increase in the variance. Interestingly, while HCP data and WS model both exhibit small world properties, their score distributions are very distinct, indicating the presence of various network topology belonging to small word regime. 

For brain connectivity models, EC and EPA have close distributions still exhibit lower values in comparison to HCP, while the DSP have high variance and a non-Gaussian behavior. 

\begin{figure}[!ht]
\includegraphics[trim={1cm 0.8cm 1.7cm 2cm},clip,width=8.6cm]{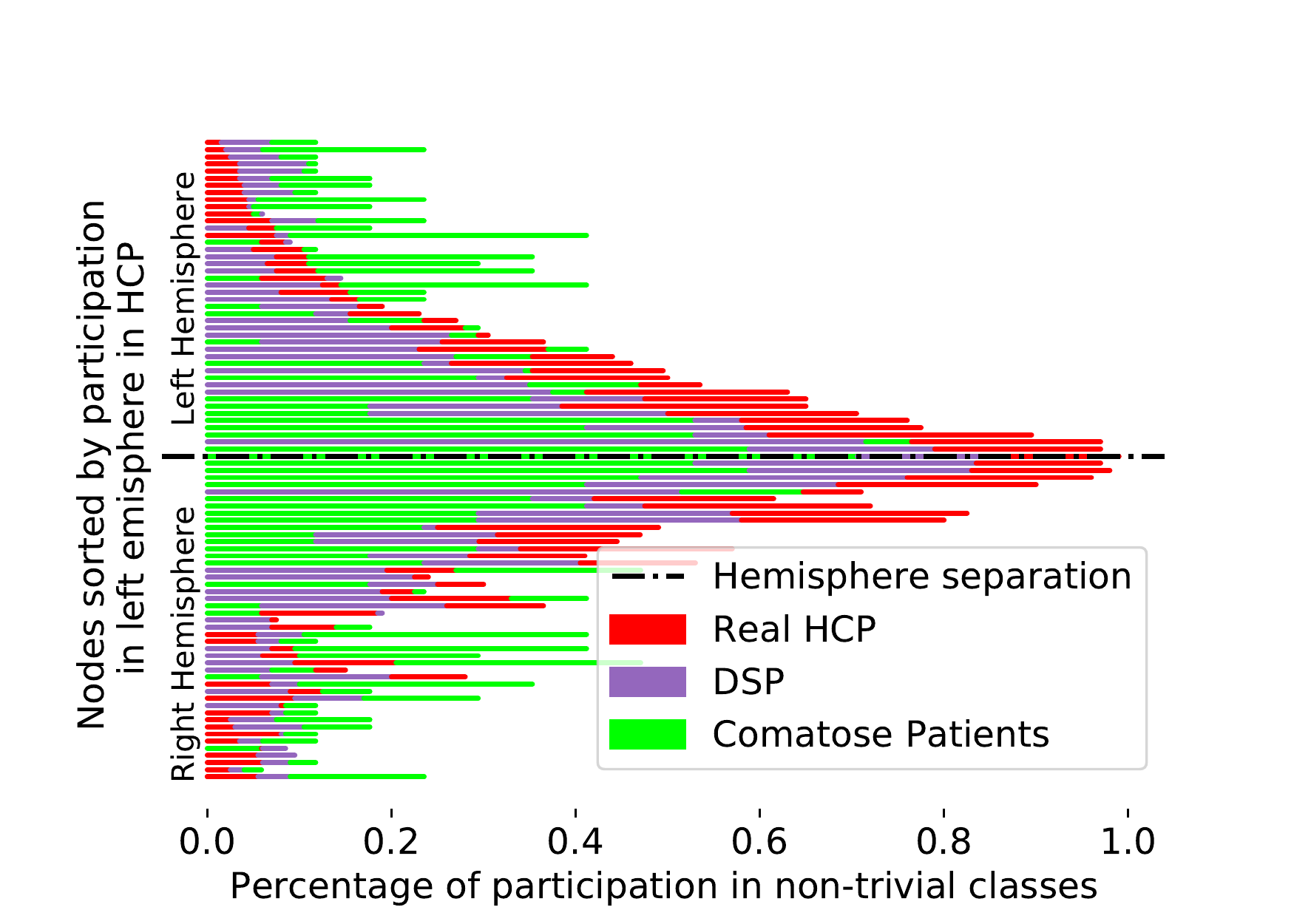}
\caption{\label{fig:participation_comatose} Nodal participation in non-trivial classes for comatose patients (n=17), healthy controls (HCP dataset) and DSP model. Left hemisphere nodes are sorted by participation in HCP, right hemispheres nodes occupy symmetrical positions of their corresponding left hemisphere nodes with respect to the dotted line.}
\end{figure}

\begin{figure}[!h]
\includegraphics[width=8.6cm]{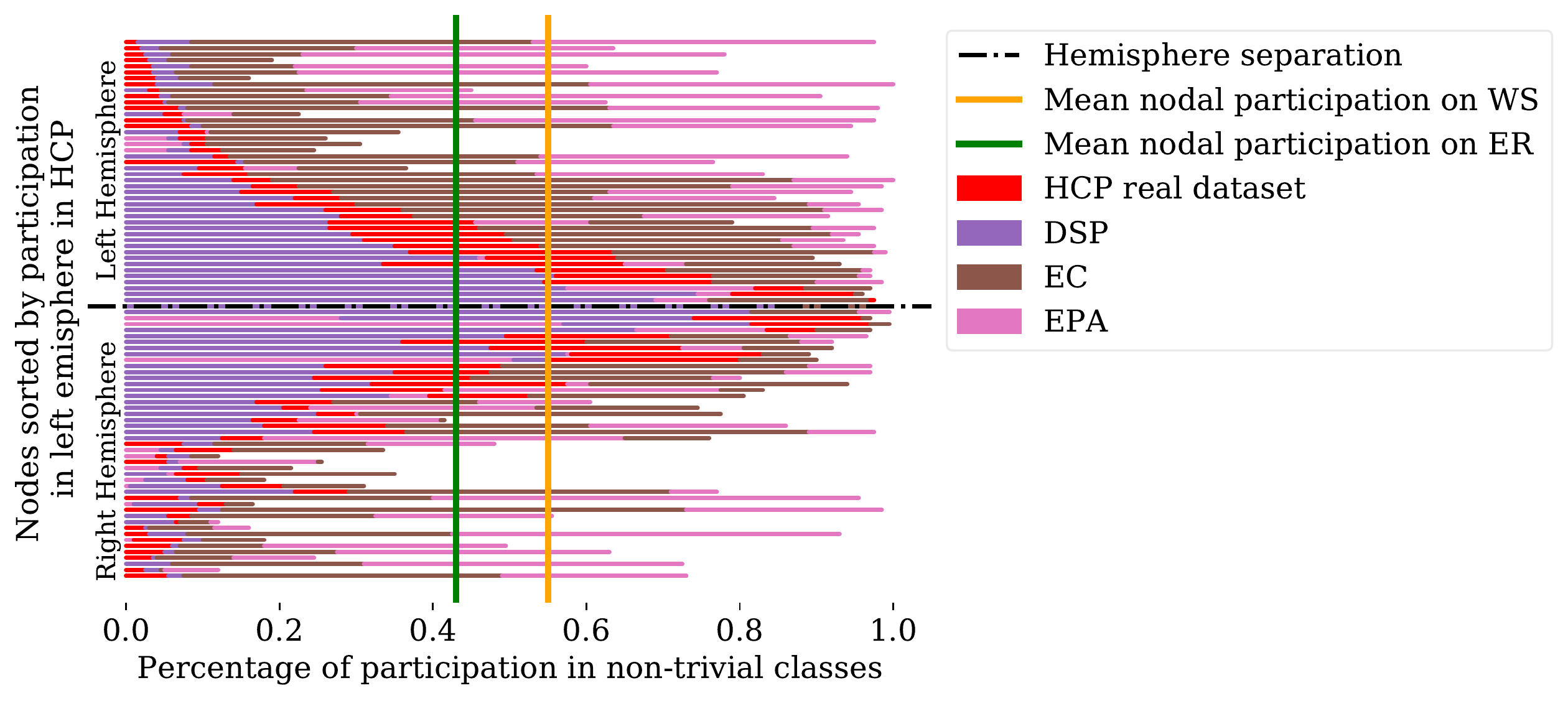}
\caption{\label{fig:nodal_participation} Nodal participation in non-trivial classes for real (HCP), and synthetic datasets. ER: Erd\H{o}s-R\'enyi model, WS: Watt-Strogatz model, DSP: Degree sequence preserving model; EPA: Economical preferential attachment model; EC: Economical clustering model.}
\end{figure}

Finally, we compare nodal participation in models and real HCP dataset (Fig. \ref{fig:participation_comatose}, SM Fig. \ref{fig:nodal_participation}). As expected, the values of percentage of participation for EC and ECA brain models, are higher than real ones, due to the spatial relations that constrain the role of each brain region. Thus, same nodes play the same role in many samples of the generated datasets. On the contrary, DSP provides a lower bound in the percentage of participation of real data. Indeed, constraining graphs to only keep same degree sequence, increases the node role variability in the group. 

A well-known brain property is the presence in the two hemispheres of homotopic regions: the right and left hemispheres are approximately mirror images of each other. That means the same region can be found in both hemispheres. An interesting result is the high number of observed classes to which belong both the homotopic regions ($38\%$ in average on HCP SM. Table  \ref{tab:homotopicalregions}). Again, when analyzing the entire data, we found that the nodal participation of brain regions is symmetrical: pairs of homotopic regions have similar percentage of participation in non-trivial classes (Fig. \ref{fig:participation_comatose}). This property is still present in the brain models that integrate the brain geometry in their construction, such as DSP (SM Fig. \ref{fig:nodal_participation}). \\
Finally, we consider 17 brain connectivity networks obtained by scanning comatose patients \cite{achard2012hubs} and we compare their node percentage of participation scores with healthy controls in Fig. \ref{fig:participation_comatose}. In comparison with HCP scores, there is less variance in the percentage participation, with almost every node close to the average percentage of $0.26\pm 0.17$ (HCP $0.32\pm 0.30$, SM Table \ref{tab:stat_part}). This makes harder to detect in comatose graphs a hierarchy in the node behavior. 
The low number of nodes sharing their property in the patient group, can be due either to the presence of many trivial classes in patient networks, or to the fact that the nodes in non-trivial classes are not consistently be the same in the group. Thus, we evaluate the number of nodes in non-trivial classes per graph, founding comparable number in controls and patients (SM Table \ref{tab:stat_trivial}). 
Thus, the difference in the percentage of participation indicates the presence of higher structural patterns variability in patients.

\section{\label{sec:6} DISCUSSION}
Graph models become largely used in real world applications and many nodal statistics have been proposed for node roles detection. However, the most informative statistics for graph comparison is highly dependent on the observed data and combining more statistics can be relevant. \\


We propose a mathematical framework with the specific purpose of providing new statistical tools for the analysis of brain functional connectivity networks. 

We introduce a nodal statistics-based equivalence relation and propose an innovative way to combine nodal statistics for graph structural pattern detection. We use the latter to compare different graphs and characterize graph family defined on the same node set. As the equivalence relation depends on a collection of nodal statistics, we define a power coefficient and an orthogonality score 
which can be used as revisited measure of nodal statistics dependency. \\

We define a graph similarity based on node roles and a mathematical tool to detect nodes persistently different from others, by computing the percentage of participation in non-trivial classes. Interestingly, nodes which tend to persistently belong to trivial class are likely to play peculiar roles in the graphs, while at the opposite nodes with a high participation, appear to share similar property with other nodes. \\

{
The proposed framework can be extended specifically to handle graph families. In order to do so, a new equivalence relation over graph instances should be introduced. In this case, the group version nodal statistics assigns a value (or an interval) to each node in the vertices-set, such as the average per node of the statistics across the graph instances (or its first-third quartile interval). Then, we introduce the corresponding nodal equivalence relation whoses node are equivalent if their average of nodal statistics is the same (or fall in the same interval). In this case, the definition of the structural pattern corresponds to \textit{an average structural pattern} of a virtual average graph. The graph family version of parsimony and orthogonality corresponds to the traditional definition on this average graph. The ability of the average structural pattern to characterize the group of graphs needs however to be explored.} \\

In terms of application, we show application in human brain functional connectivity networks. We report high correspondence scores among networks of healthy controls and differences in the nodal participation in non-trivial classes of random and real graphs. Interestingly, NPP can detect brain homotopy. Applied to comatose patients, our mathematical framework allows to quantify at the global and nodal levels how their functional connectivity networks differ from healthy controls. These results motivate further investigations, in particular for a deeper characterization of each identified class. {For instance, with a counting of nodes not only on trivial classes, but on different class sizes.} \\


\section{Acknowledgments}
L. Carboni is the recipient of a grant from MIAI@Grenoble Alpes (ANR 19-P3IA-003).

\bibliography{sorsamp}

\section*{\label{sec:SuppMat} Supplementary Material}

\begin{figure}[!ht]
\includegraphics[trim={4.5cm 16.7cm 4cm 3.5cm},clip, width=8.6cm]{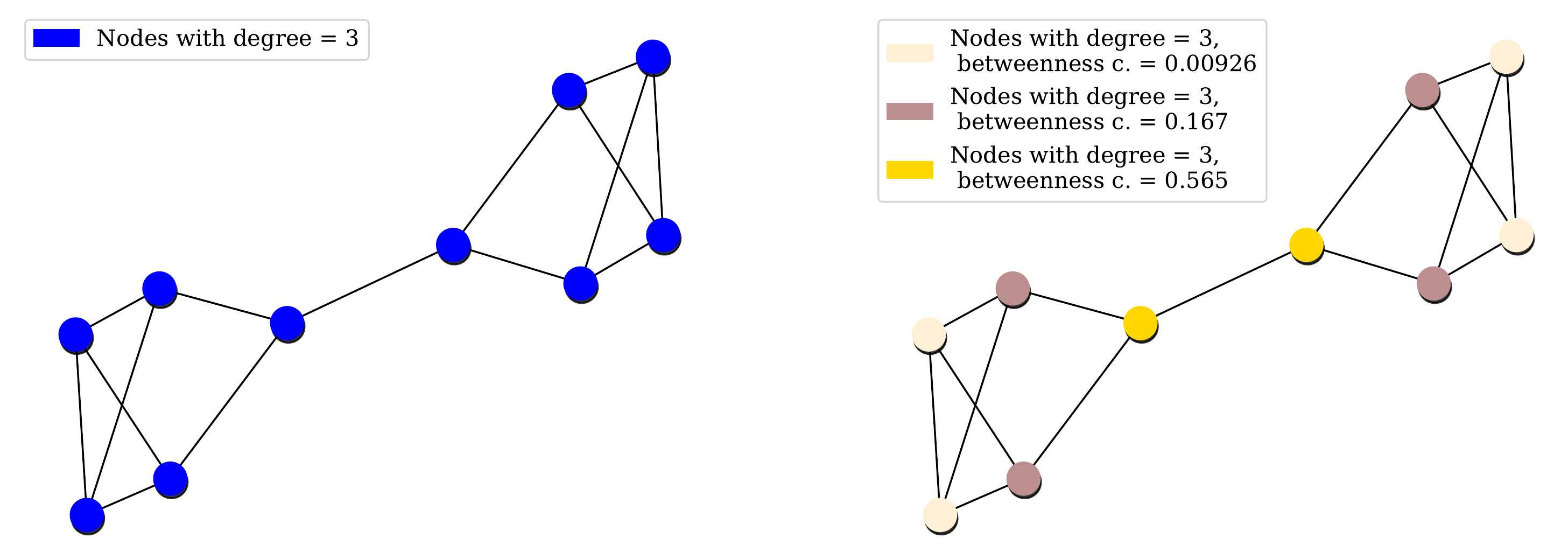}
\caption{\label{fig:ORBIT} Visualization of the structural pattern associated with a given statistics on a trivial graph whose nodes have the same degree. Left: Structural pattern associated with nodal degree. Right: Structural pattern associated with combination of betweenness centrality and degree. Colors correspond to different classes. In this toy example, the degree alone is not sufficient to reveal different equivalence classes and identifies a unique class. While, when two nodal statistics are considered, a non-trivial structural pattern appears.}
\end{figure}

\subsection{General properties}
\subsubsection{Properties of PC}
\begin{figure}[!hb]
\includegraphics[width=8.4cm]{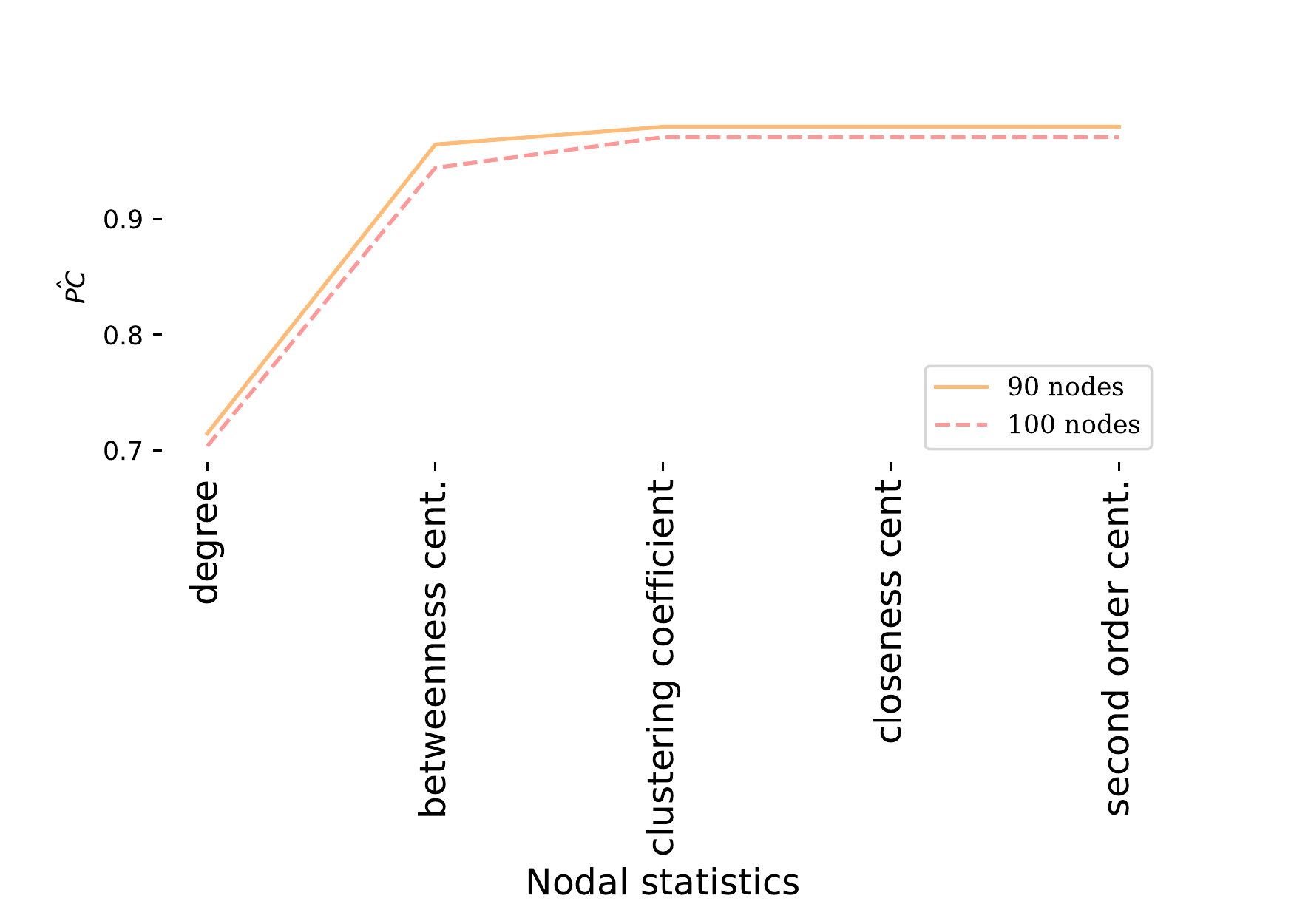}
\caption{\label{fig:pc} Normalized power coefficient ($\hat{\text{PC}}$) on nodal statistics incremental sets of two Erd\H{o}s-R\'enyi graphs of 90 and 100 nodes.}
\end{figure}

{Note that all the listed properties are true also for $\widehat{\text{PC}}$.}
\begin{itemize}
    \item[.] on the same graph the PC increases on increasing collections of nodal statistics (SM Fig. \ref{fig:pc});
    \item[.] the PC of every nodal statistics collection equals zero for vertex-transitive graph;
    \item[.] if the PC of a nodal statistics collection equals the PC of one of its element, then the correspondence structural patterns score of the structural pattern associated with the collection and the one of that element is 1;
    \item[.] if the PC of a graph equals $0$ for one collection of statistics, then the graph does not admit non-trivial automorphisms;
    \item[.] if two graphs are isomorphic than their PC is the same for all statistics collection.
\end{itemize}
{Relation with network ensembles entropy.\\
The number of eligible automorphisms of a graph corresponds to the number of rows permutations of its adjacency matrix. Following \cite{bogacz2006homogeneous}, the partition function of the ensemble of a given topology $ \mathcal{G} = (\mathcal{V},\mathcal{E}) $, with $\text{Aut}(\mathcal{G})$ the set of automorphism of $\mathcal{G}$
\begin{equation}
    Z( \mathcal{G}=(\mathcal{V},\mathcal{E}) ) = \Big | \frac{|\mathcal{V}|!}{|\text{Aut}(\mathcal{G})|} \Big | .
\end{equation}
We denote $\text{PC}^*$ the PC computed for a collection of statistics whose equivalence relation corresponds to the automorphisms relation.
Then, we have 
\begin{eqnarray}
\text{PC}^* = |\log \frac{1}{\text{Z}}| \\
\text{PC}^* = | - \log {\text{Z}}| \\
\text{PC}^* = \log {\text{Z}} \\
 \text{entropy} \propto \text{PC}^*
\end{eqnarray}  
This is in line with the idea that a higher level of order in the graph structure is associated with lower entropy, indeed for vertex-transitive graphs the entropy reaches is minimal value of zero \cite{bianconi2007entropy,bianconi2009entropy}. A comparison with-in entropy and PC$^{*}$ can be found in Table \ref{tbl:entropy}. Note that in the first and last examples the statistics collection choice does not affect the PC.  
}

\begin{table*}[h!]
\caption{Entropy and PC$^*$ on known graphs}

  \begin{tabular}{| c | c | c | }
    \hline
    Graph & Entropy & PC$^*$ \\ 
    \hline
    \begin{minipage}{.3\textwidth}
      \includegraphics[width=\textwidth]{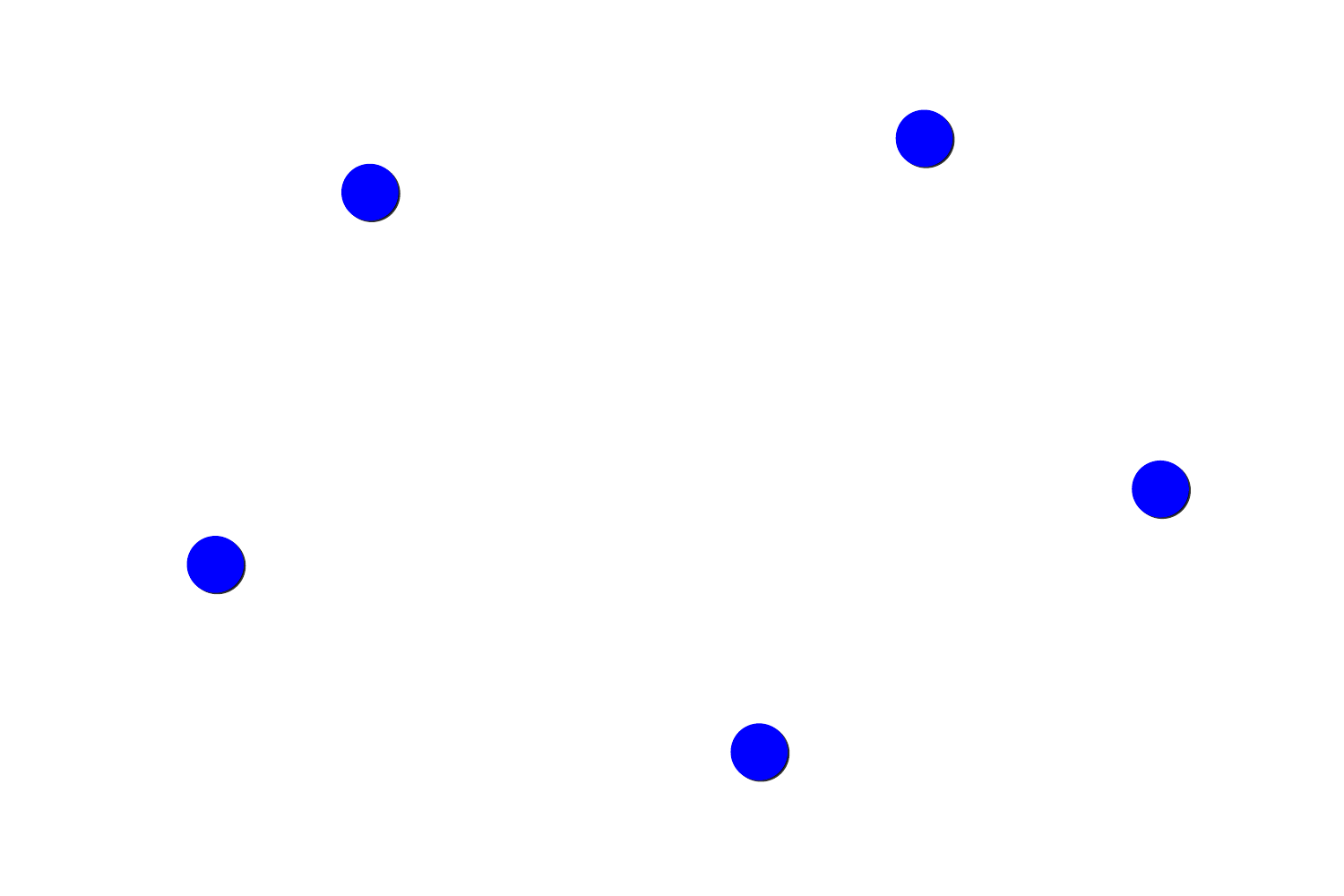}
     Trivial graph with $n$ isolated nodes\\
    \end{minipage}
    &
    \begin{minipage}{.3\textwidth}
    \begin{equation*}
        \frac{1}{n}\log{\frac{n(n-1)}{2} \choose 0} = 0
    \end{equation*}   
    \end{minipage}

    & 
    \begin{minipage}{.3\textwidth}
    \begin{eqnarray*}
    {\forall \mathcal{S}}, 
     \quad \text{PC}(\mathcal{S}) &=& \text{PC}^* \\
    \text{PC}^* &=&\log \frac{n!}{n!} = 0 \\
    \end{eqnarray*}   
    \end{minipage}
    \\
\hline
 \begin{minipage}{.3\textwidth}
      \includegraphics[width=\textwidth]{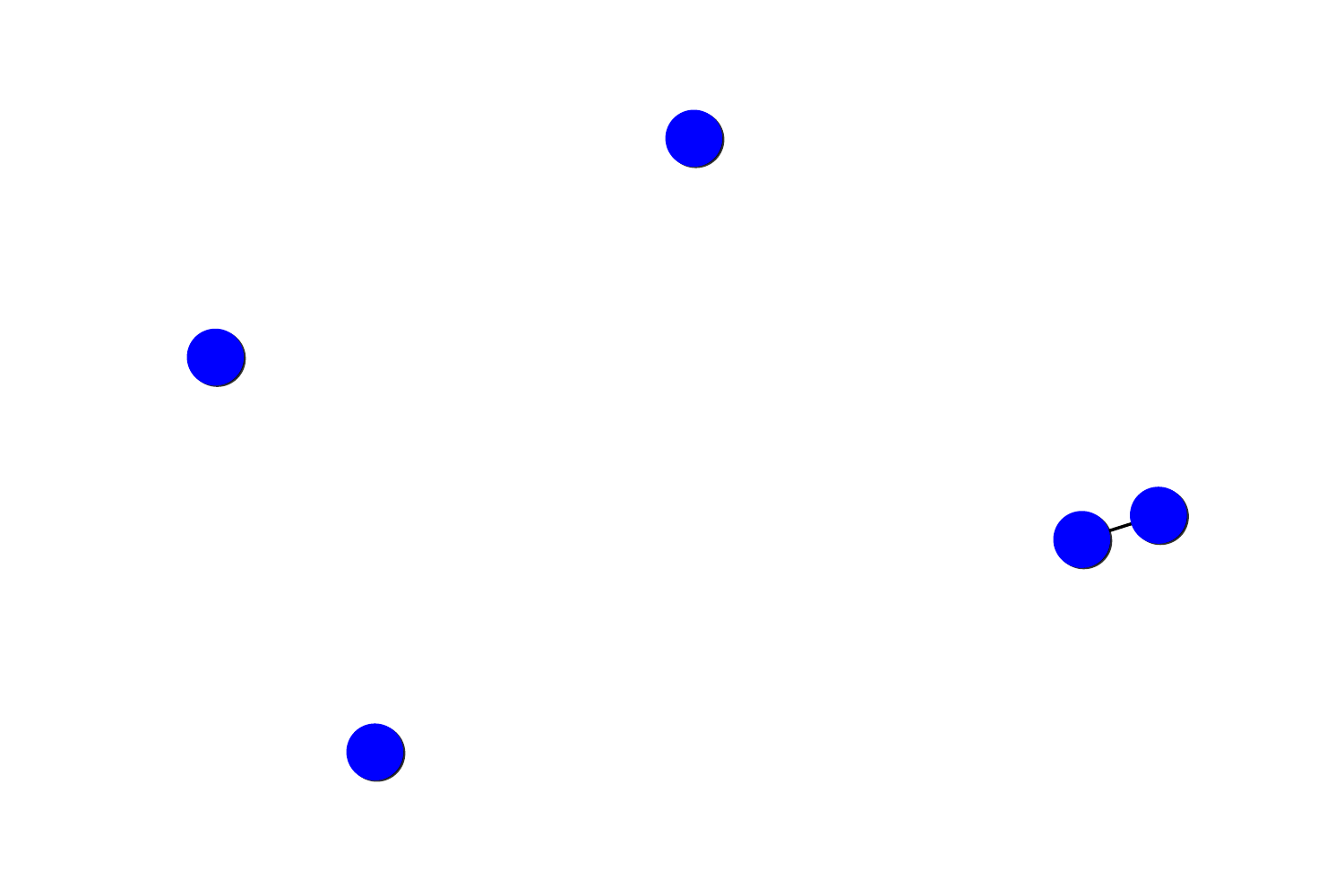}
     ER model $p{n \choose 2} = 1$\\
     2 connected nodes and $n-2$ isolated nodes 
     \vspace{1mm}
    \end{minipage}
    &
    \begin{minipage}{.3\textwidth}
    \begin{equation*}
       \frac{1}{n}\log{\frac{n(n-1)}{2} \choose 1} = \frac{1}{n}\log \frac{n(n-1)}{2} 
    \end{equation*}   
    \end{minipage}

    & 
    \begin{minipage}{.32\textwidth}
    \begin{eqnarray*}
    {\mathcal{S}}=
    \{\deg\} \quad \text{PC}(\mathcal{S}) &=& \text{PC}^*\\
    \text{PC}^* &=&\log \frac{(n-2)!2}{n!} = \\
    &=&\log \frac{2}{n(n-1)} 
    \end{eqnarray*}   
    \end{minipage}\\
\hline
 \begin{minipage}{.3\textwidth}
      \includegraphics[width=\textwidth]{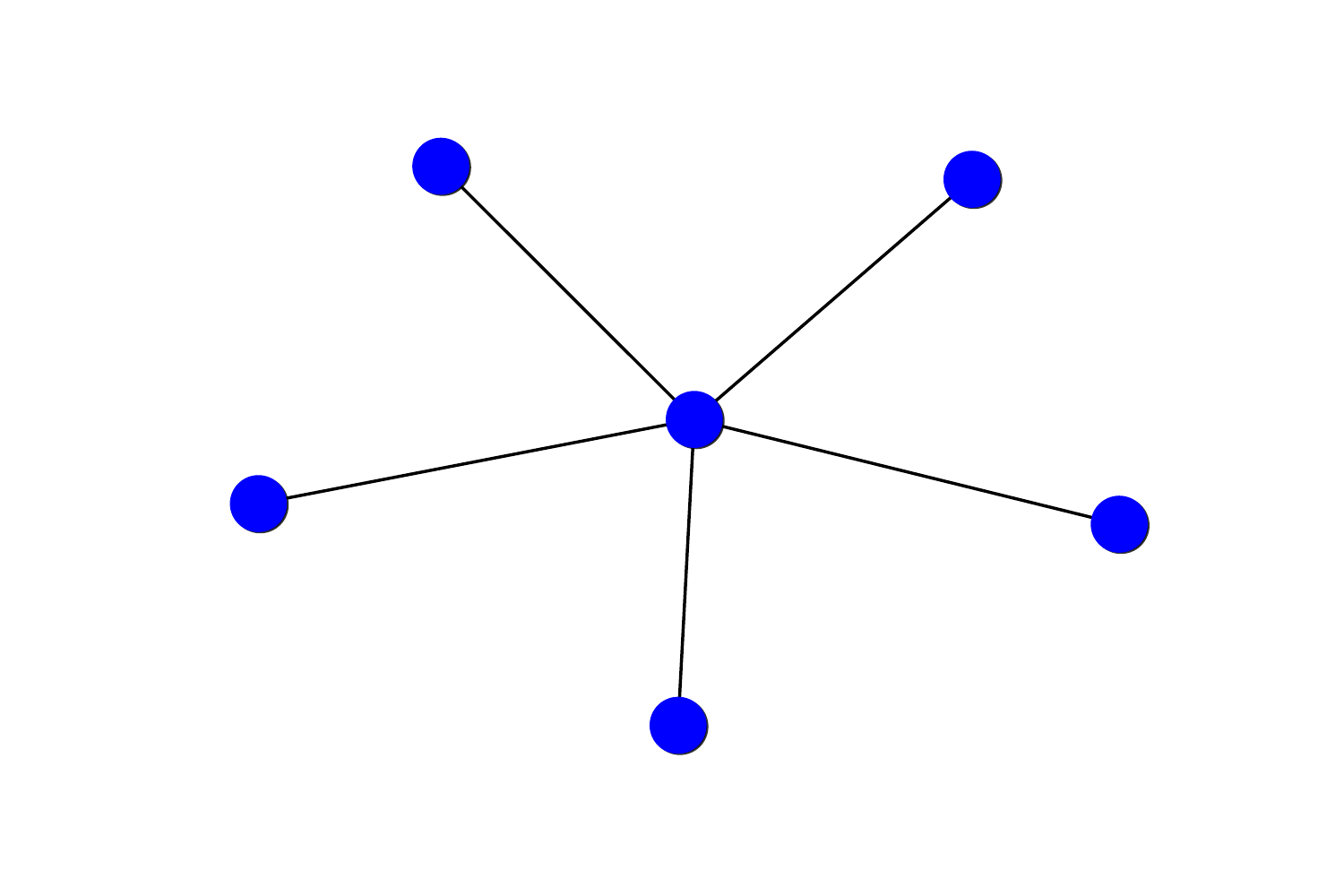}
     Star graph with $n$ nodes and $n-1$ edges 
    \end{minipage}
    &
    \begin{minipage}{.3\textwidth}
    \begin{equation*}
        \frac{1}{n} \log \frac{n!}{(n-1)!} = \log n
    \end{equation*}   
    \end{minipage}
    & 
    \begin{minipage}{.3\textwidth}
    \begin{eqnarray*}
    {\mathcal{S}}=
    \{\deg\} \quad
    \text{PC}(\mathcal{S}) &=& \text{PC}^*\\
    \text{PC}^* &=&\log \frac{(n-1)!}{n!} =  \\
    &=&\log \frac{1}{n} 
    \end{eqnarray*}   
    \end{minipage}\\
\hline
\begin{minipage}{.3\textwidth}
      \includegraphics[width=\textwidth]{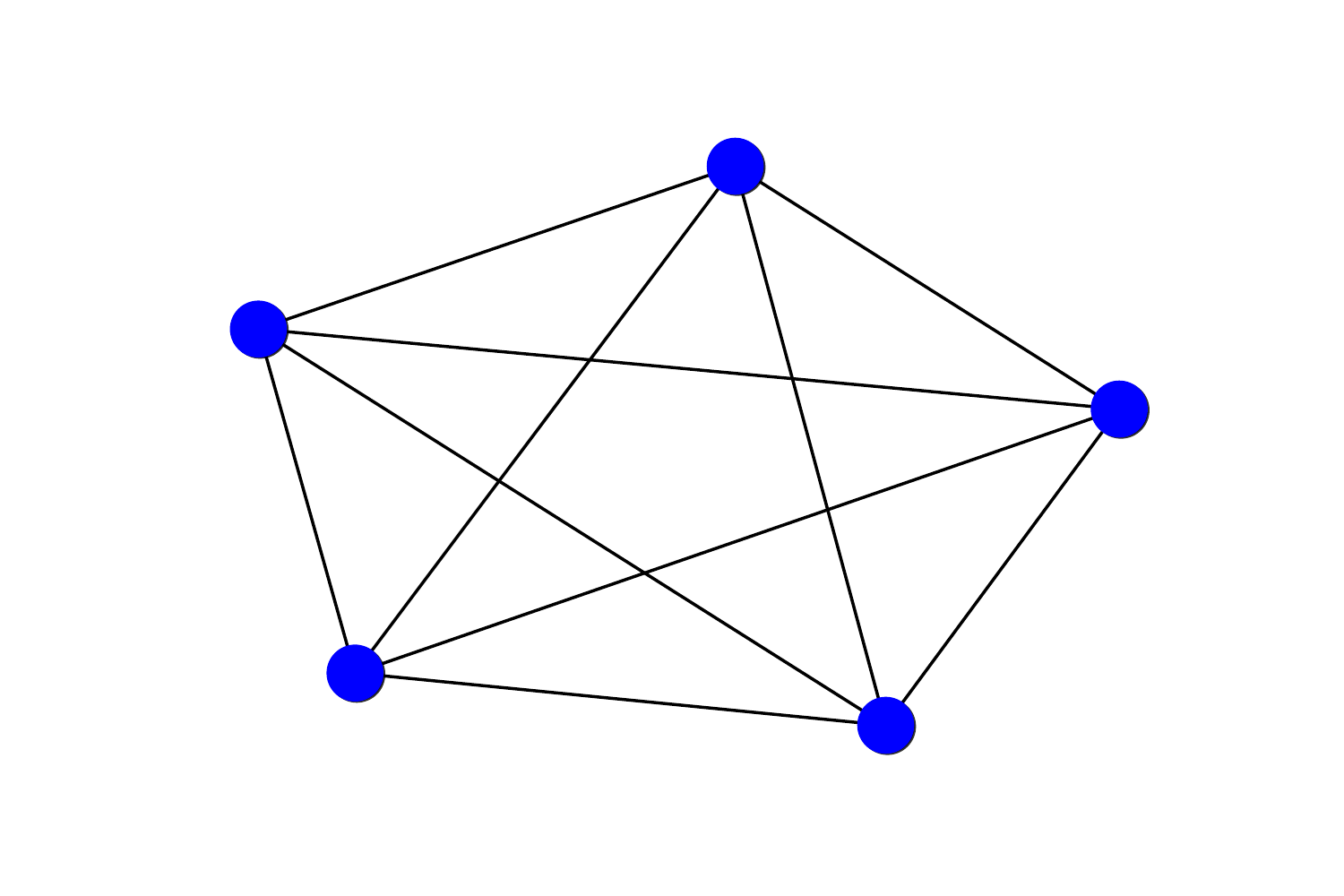}
     Complete graph with $n$ nodes having $n-1$ edges 
    \end{minipage}
    &
    \begin{minipage}{.3\textwidth}
    \begin{equation*}
        \frac{1}{n} \log \frac{n!}{n!} = 0
    \end{equation*}   
    \end{minipage}
    & 
    \begin{minipage}{.3\textwidth}
    \begin{eqnarray*}
    \forall \mathcal{S}, \quad
    \text{PC}(\mathcal{S}) &=& \text{PC}^*\\
    \text{PC}^* &=&\log \frac{n!}{n!} =  \\
    &=& 0  
    \end{eqnarray*}   
    \end{minipage}\\
\hline
  \end{tabular}
  \label{tbl:entropy}
\end{table*}


\subsubsection{Properties of Orthogonality}
\begin{itemize}
    \item a nodal statistics whose induced partition is composed of classes having each one a unique element, is perfectly orthogonal with every nodal statistics;
    \item if collection of statistics is perfectly orthogonal, all other collection having as subset that collection is perfectly orthogonal as well;
    \item if a perfectly orthogonal statistics set exists on a graph, then the graph does not admit non-trivial automorphisms.
\end{itemize}

\subsubsection{Properties of Correspondence of structural pattern}
\begin{itemize}
    \item[.] all graphs defined on the same node set, having same degree sequence, have a correspondence of structural patterns associated with the degree statistics equals to 1 ;
    \item[.] the minimum values of structural pattern score is given by $\frac{1}{|\mathcal{V}|}$;
    \item[.] if on the same graph, the structural patterns score of different nodal statistics reaches the minimal value, then the nodal statistics are perfectly orthogonal. 
\end{itemize}
\newpage

\subsection{Generative Networks}
\begin{figure*}[!ht]
    \centering
    \includegraphics[trim={10cm 0 1cm 0},clip,width=17.2cm]{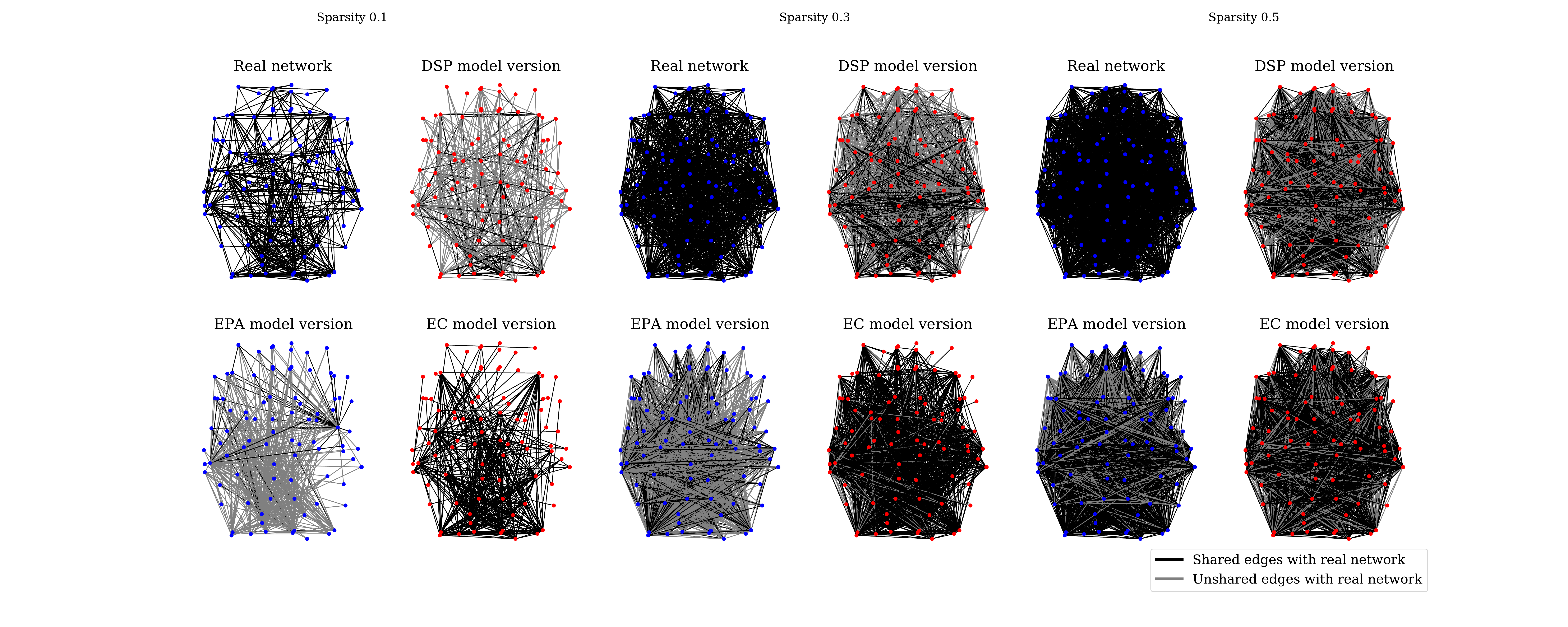}
    \caption{\label{fig:brainvisualization} Examples of real functional connectivity network in HCP dataset and the corresponding model versions for different sparsity values. DSP: Degree sequence preserving model; EPA: Economical preferential attachment model; EC: Economical clustering model.}
\end{figure*}

\subsubsection{Erd\H{o}s-R\'enyi model (ER)} 
The Erd\H{o}s-R\'enyi (ER) model generates a binomial graph $\mathcal{G}_{n,p}$ by the creation of edges among $n$ nodes. Each edge has a probability $p$ of being created. The expected number of edges in  $\mathcal{G}_{n,p}$ is then $p {n\choose 2}$ and its sparsity ration equals $p$. For values of $p$ close to 1, the graph tends to be the complete graph in which all possible edges are present. 
\subsubsection{Watts-Strogatz model (WS)}
The Watts-Strogatz (WS) model generates a small-world graph $\mathcal{G}_ {n,k,p}$ by connecting each node with its $k$ neighbors nodes and then recombining each edge with probability $p$. In this case, the number of created edges is always $\frac{nk}{2}$, requiring an even value for $k$, which corresponds to a sparsity value of  $\frac{nk }{2{n\choose 2}}=\frac{k}{(n-1)}$. The $p$ parameter, which regulates the probability of rewiring the edges, generates the regular graph ($p=0$) in which all nodes have the same degree, and the completely random graph ($p=1$) in which the expected number of edges are randomly distributed on the vertices set. We consider cases $p=0.1,0.5, 0.9$ and refer to the case $p=0.5$ as the small-world model. 
\subsubsection{Barab\'asi-Albert model (BA)}
The Barab\'asi-Albert (BA) model generates a graph $\mathcal{G}_ {n,m}$ by favoring specific attachments. It starts from a star graph of $m+1$ nodes and attaches the $n-m-1$ remaining nodes to the $m$ existing nodes with high degree. In that case, the number of edges expected is given by the sum of the first $m$ edges of the initial graph with $(n-m-1) m$ edges created by attaching new nodes until the graph has $n$ vertices leading to the sparsity value equals to $\frac{m(n-m)}{{n\choose 2}}$. In this case, having fixed $n$ and the level of sparsity $l$, there are two possible choices for $m$, corresponding to the solutions of $$m^2-mn+l{n\choose 2}=0 .$$ The existence of real solutions to the previous equations is only guaranteed for $ l \leq \frac{n^2}{4{n\choose 2}}$ and in that case, it always has two positive solutions.
We considered both cases, referring to BA1 and BA2, respectively for the lower and the highest root. Due to the constraints of existence of real solutions, all networks generated according to Barab\'asi-Albert model are sparse \cite{del2011all}. 
\subsubsection{Degree sequence preserving model (DSP)} 
The degree sequence preserving (DSP) model is based on  the configuration model \cite{barabasi1999emergence}. For each graph from our real dataset (HCP), we search for preserving its degree sequence while controlling the sparsity ratio. For this reason, given the correlation matrix associated with a subject and given a sparsity ratio, we threshold the correlation matrix to obtain a binary version with the number of edges corresponding to the fixed sparsity. Then, we extract the degree sequence and randomly generate a new graph that preserves the given degree sequence. Since the degree of each node is fixed, we obtain a synthetic graph which has the same sparsity as its real version. In such a way, for all sparsity values we considered, we obtain the synthetic graphs whose elements are the \textit{model version} of the corresponding real graphs. An example of the simulated DSP networks is shown in SM Fig. \ref{fig:brainvisualization}. 

\subsubsection{Economical preferential attachment model (EPA)}
The economical preferential attachment (EPA) model has been defined to reproduce functional brain networks \cite{vertes2012simple}. The probability of observing a connection between region $i$ and region $j$ is given by $$p_{i,j} \propto \big(\deg(i)\deg(j)\big)^{\gamma} (d_{i,j})^{- \eta} $$
where $\deg(i)$ is the degree of node $i$ and $d_{i,j}$ is the Euclidean distance in anatomical space between $i$ and $j$. Since we want to generate network at fixed sparsity, given a real network, we first extract its degree distribution. Next, we compute the $p_{i,j}$ of all possible pairs of nodes and then we select the highest probability until we reach the expected number of edges. To ensure connectivity, we also add the Minimum Spanning Tree as it is done in real data. The parameters $\gamma,\eta$ are tuned according to \cite{vertes2012simple} and fixed to respectively $1.81$ and $5.37$. An example of the simulated EPA networks is shown in SM Fig. \ref{fig:brainvisualization}. 

\subsubsection{Economical clustering model (EC)} 
The economical clustering (EC) model has also be proposed in the context of functional brain networks \cite{vertes2012simple}. The probability of observing a connection between region $i$ and region $j$ is given by $$p_{i,j} \propto (k_{i,j})^{\gamma} (d_{i,j})^{- \eta} $$
where $k_{i,j}$ is the number of nearest neighbors in common between nodes $i$ and $j$, while $d_{i,j}$ is the Euclidean distance in anatomical space between $i$ and $j$. For being able of tuning the sparsity of the model, we generate an EC model version of real network. Given a real network at a given sparsity ratio, we determine its $k_{i,j}$ and compute the $p_{i,j}$ of all possible node pairs. Finally, we select edges whose probability is higher until the expected number of edges is reached. Again, we ensure connectivity by adding missing edges from the Minimum Spanning Tree algorithm. 
The parameters $\gamma$, and $\eta$ are fixed to $3.17$ and  $2.63$ respectively. For these values the model best fits data both on training and validation set \cite{vertes2012simple}. An example of the simulated EC networks is shown in SM: Fig. \ref{fig:brainvisualization}.

\subsection{More experiment results}
~

~
\begin{table}[!ht]
\caption{\label{tab:stat_trivial} Nodes in non-trivial class per graph}
\begin{tabular}{ ccccc } 
 & AVG & SD & MIN  & MAX \\ 
\hline
 HCP & $0.32$ & $0.062$ & $0.17$ & $0.52$ \\ 
Comatose & $0.26$ &  $0.067$ & $0.13$ & $0.41$ \\ 
\end{tabular}
\end{table}
\begin{table}[!ht]
\caption{\label{tab:stat_part} Statistics on nodal percentage of participation }
\begin{tabular}{ ccccc } 
 & MIN & MAX & AVG & SD \\  \hline  HCP & $0.015$  & $0.98$ & $0.32 $ & $0.30$  \\   Comatose & $0.0$ &
 $0.76$ & $0.26$ & $0.17$\\ 
\end{tabular}
\end{table}
\begin{table}[!ht]
\caption{\label{tab:homotopicalregions} Ratio of nodes in non-trivial classes in the same class of their homotopical in HCP dataset.}
\begin{tabular}{ ccccc } 
 & AVG & SD & MIN  & MAX \\ 
\hline
 & $0.38$ & $0.11$ & $0.13$ & $0.67$ \\ 
\end{tabular}
\end{table}
\begin{figure}[!htp]
\includegraphics[width=8.6cm]{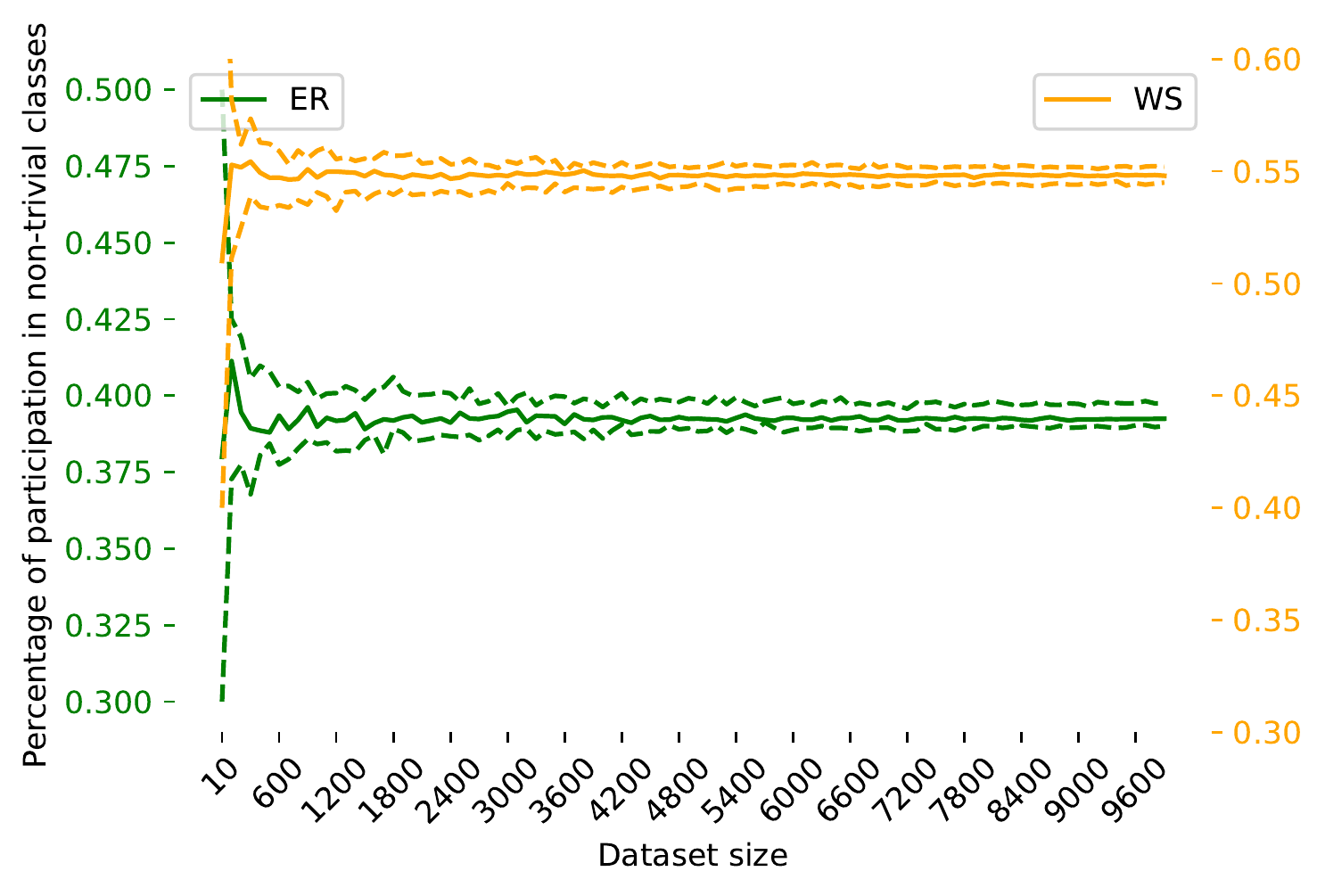}
\caption{\label{fig:mean_part} Mean participation of a node in datasets of different size. Mean percentage of participation in non-trivial classes for a single node in 90 nodes ER  (Erd\H{o}s-R\'enyi) and WS (Watt-Strogatz) model at 0.1 sparsity level. The results indicate that the mean percentage of participation stabilizes respectively at $0.43$ and $0.55$ for ER and WS models. WS is obtained with 0.5 edge rewiring probability.  
Dots lines are first and third quartile. }
\end{figure}
~
\newpage
~
\newpage
~
\newpage
\begin{figure}[!h]
\includegraphics[width=8.6cm]{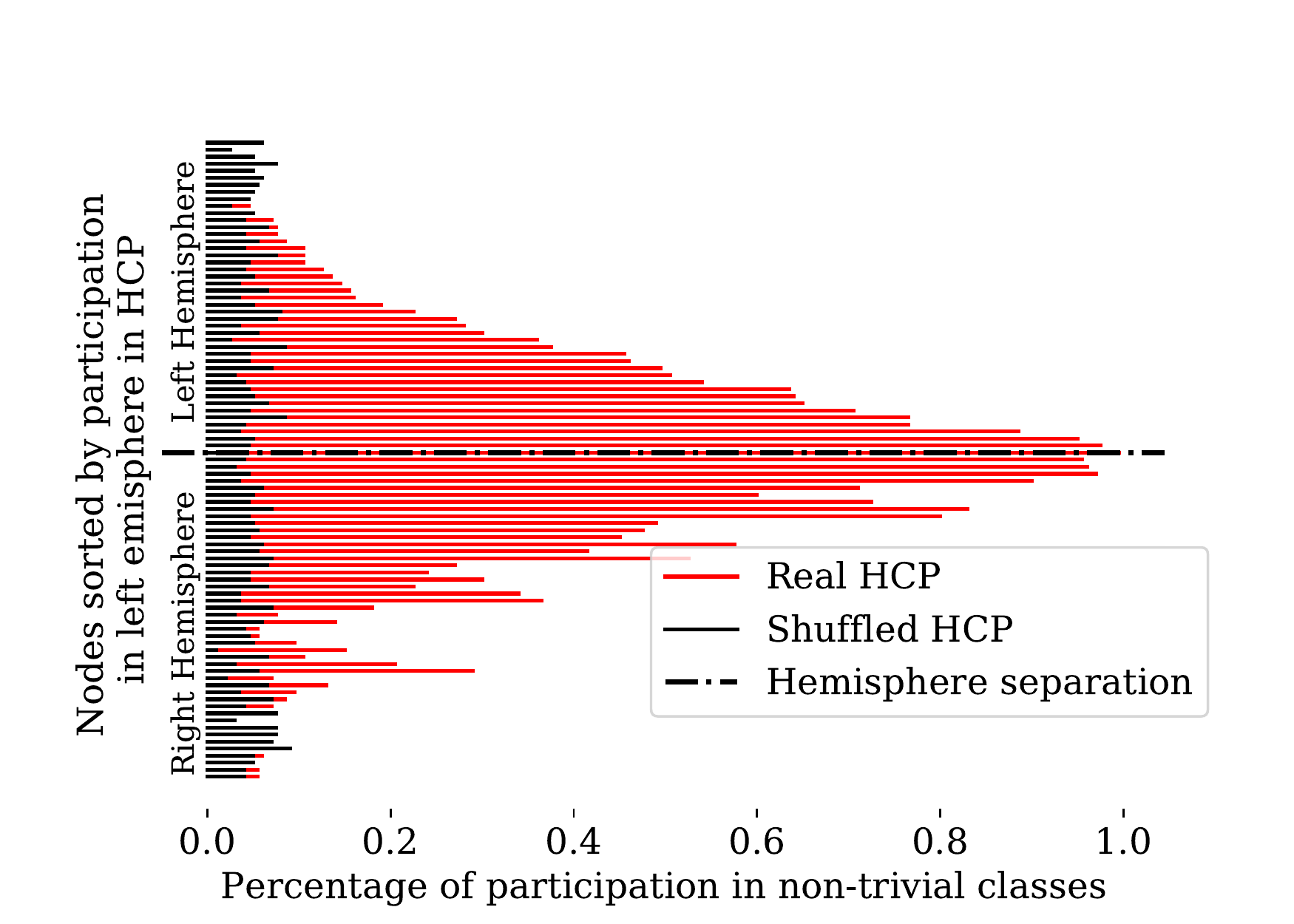}
\caption{\label{fig:nodal_participation_hemisphere} Nodal participation in non-trivial class in HCP dataset and a shuffled version. Nodes labels are sorted according to the percentage of participation of left hemisphere regions. The symmetry reveals the expected hemisphere similarity in the participation of analogue regions.The percentage of participation of each node is also compared with a shuffled HCP dataset, where each real network is re-ordered by a random shuffle of the adjacency matrix, preserving the degree distribution. In this way, we expect that non-zero percentage of participation is simply due to chance. The participation indices for this random dataset appear to be significantly lower than the ones observed in the real HCP. However, even if closer to 0, all nodes appear to participate at least in one non-trivial class. Thus, when, for the real data, we observe a high participation index, we can conclude that the node is likely to share its equivalence role with some other nodes in the graph. At the same time, when a node does not have a positive percentage of participation, we expect the node to be unique, consistently in all networks and so to retrieve regions associated with unique functions. }
\end{figure}
\end{document}